\documentclass[preprintnumbers,amsmath,amssymb,floatfix,11pt,prd,onecolumn,superscriptaddress,nofootinbib]{revtex4}
\usepackage[utf8]{inputenc}
\usepackage{diagbox}
\usepackage{latexsym}
\usepackage{epsfig}
\usepackage{epstopdf,float}
\usepackage{graphicx}
\usepackage{amssymb}
\usepackage{amsmath}
\usepackage{dcolumn}
\usepackage{bm}
\usepackage{color}
\usepackage{comment}
\usepackage[shortlabels]{enumitem}
\usepackage{subfigure}
\usepackage{multirow}
\usepackage{xcolor}
\begin{document}
\title{\bf Imprints of Black Hole Shadows and Polarization Patterns of Various Thick Disks: Bumblebee gravity}
\author{Muhammad Israr Aslam}
\altaffiliation{mrisraraslam@gmail.com,
israr.aslam@umt.edu.pk}\affiliation{Department of Mathematics,
School of Science, University of Management and Technology,
Lahore-$54770$, Pakistan.}
\author{Nazek Alessa}\altaffiliation{naalessa@pnu.edu.sa}\affiliation{Department
of Mathematical Sciences, College of Science, Princess Nourah bint
Abdulrahman University, P.O.Box 84428, Riyadh 11671, Saudi Arabia}
\author{Chen-Yu Yang}
\altaffiliation{chenyu\_yang2024@163.com} \affiliation{Department of
Mechanics, Chongqing Jiaotong University, Chongqing 4000, People's
Republic of China}
\author{Xiao-Xiong Zeng}
\altaffiliation{xxzengphysics@163.com (Corresponding author)}
\affiliation{College of Physics and Optoelectronic Engineering,
Chongqing Normal University, Chongqing $401331$, People's Republic
of China}


\begin{abstract}
The main objective of this study is to explore the shadow and
polarization patterns of a Kerr-Sen-like BH induced from Bumblebee
gravity, which, among other alternative theories of gravity beyond
Einstein gravity, stands out as a promising candidate for explaining
certain high-energy astrophysical phenomena. Specifically, we would
like to probe the influence of the rate of LSB parameter $\ell$ and
the Bumblebee charge $Q$ on the resulting image morphology at
$230\mathrm{GHz}$. We adopt a phenomenological RIAF-like model and
an analytical BAAF disk model. Both models depict that the bright
ring is encircled by two central dark regions, each of which
gradually shrinks with increasing $\ell$. Consequently,
frame-dragging gives rise to a pronounced brightness asymmetry,
which is more enhanced with increasing $Q$. A notable feature in the
anisotropic emission case is the emergence of a vertically
stretched, elliptical ring structure. Compared with the RIAF
framework, the bright ring in the BAAF disk images appears
geometrically thinner, and the separation between the primary and
higher-order images becomes more pronounced. Finally, the
polarization patterns trace the brightness distribution and vary
with both $\ell$ and $Q$, reflecting the spacetime structure. These
results demonstrate that intensity and polarization in thick disk
models provide probes of Kerr-Sen-like BHs and near-horizon
accretion physics
\end{abstract}
\date{\today}
\maketitle

\section{Introduction}
Recent experimental development in the fabric of gravitational
physics has ushered in a new window in which electromagnetic
radiation is used to probe the nature of ultra-compact objects,
including BHs. Among the most significant breakthroughs are the
discoveries by the Event Horizon Telescope (EHT), which imaged the
electromagnetic emission from super-heated plasma orbiting the
supermassive objects at the center of both the M$87$ and Milky Way
galaxies \cite{bm1,bm2,bm3,bm4}. Complementing these achievements,
the GRAVITY instrument has detected infrared flares in the immediate
vicinity of our Galactic center \cite{bm5,bm6}. This emerging
observational window enables the gravitational physics community to
rigorously test theoretical models through the imaging of BHs,
particularly by comparing observations with General Relativistic
Magneto-Hydrodynamic (GRMHD) simulations of the surrounding plasma
\cite{bm7,bm8}.

The solution of Kerr BH serves as a fundamental cornerstone for
several fundamental theorems within the framework of general
relativity (GR), which establish it as the unique stationary and
axisymmetric vacuum solution of the Einstein field equations endowed
with Killing horizons \cite{bm9}. Moreover, recent EHT experimental
achievements are consistent with the theoretical predictions of the
Kerr hypothesis \cite{bm10,bm11}. Based on this hypothesis, the
complete gravitational collapse of a body in a realistic
astrophysical setting leads to the formation of a rotating,
electrically neutral BH \cite{bm12,bm13}. Consequently, the Kerr
paradigm has emerged as the standard mechanism against which
observational data are interpreted, while alternative space-times
describing exotic compact objects must demonstrate superior
agreement with observations to be considered viable
\cite{bm14,bm15}. In this perspectives, despite the good agreement
of GR with observational constraints, this theory still faces
several significant challenges, specifically when integrated with
the Standard Model of particle physics, leading to the pursuit of
modified theories of gravity. These challenges leave open the room
of probing new physics beyond GR within the limits of current
precision. Among various extensions, Lorentz symmetry breaking (LSB)
has emerged as a promising window into the potential quantum nature
of gravity at the Planck scale \cite{bm16,bm17}.

Although Lorentz symmetry serves as a fundamental cornerstone of
both quantum field theory and GR, its exact validity across all
energy scales remains an open question. Many exciting theories of
quantum gravity suggest that this symmetry may be violated at
sufficiently high-energy scales \cite{bm18,bm19}, potentially giving
rise to observable signatures in low-energy effective descriptions
\cite{bm20}. In this regard, the authors in \cite{bm21} construct
the Standard Model Extension mechanism, to systematically
characterize the LSB effects, which provide a unified framework for
describing Lorentz violation. From the gravity perspectives, the
Einstein-Bumblebee model stands as one of the simplest
implementations of the Standard Model Extension
\cite{bm18,bm22,bm23}. It induces spontaneous LSB by defining a
vector field with a non-zero vacuum expectation value. In this
scenario, using the mechanism of Einstein-Bumblebee gravity model,
many BH solutions have been derived, for instance one can see Refs.
\cite{bm24,bm25,bm26,bm27,bm28}. Among them, a Kerr-Sen-like BH
solution \footnote{A charged and rotating BH solution was proposed
under the scheme of heterotic string theory, which is well-known as
the Kerr-Sen BH \cite{bm29}.} has been obtained from the
Einstein-Bumblebee action, in which two extra parameters,
corresponding to the rate of Lorentz symmetry breaking $L$ and the
charge in Kerr-Sen BH $Q$, are spontaneously added
\cite{bm24,bm27,bm30,bm31}. The observational constraints of these
two parameters were carefully investigated from the observations of
supermassive BHs \cite{bm32,bm33,bm34,bm35,bm36}. More recently,
several significant properties of BH solutions have been explored in
the context of Bumblebee gravity model such as, energy extraction
via magnetic reconnection \cite{bm37}, shadow observables
\cite{bm38}, vacuum solutions \cite{bm39}, thermodynamics
\cite{bm40} dark matter spike \cite{bm41} and the process of
accretion disk \cite{bm42} etc.

The BH shadow is a dark region formed by the deflection of light in
the strong gravitational field of a BH. Its morphology and radius
are primarily evaluated by the underlying space-time geometry,
allowing one to infer the significant properties of BHs with the
analysis of shadow dynamics and its associated consequences.
Continued advancements in experimental physics and observational
astronomy have significantly deepened our understanding of BHs. It
is widely acknowledged that when matter is captured by the
gravitational field of a rotating supermassive BH, frame-dragging
effects compel the particles to co-rotate with the BH. Subsequently,
the infalling matter forms a hot, magnetized plasma that radiate
thermal synchrotron radiation, producing a luminous accretion disk.
And for rapidly spinning BHs, electromagnetic energy extraction
mechanisms can further drive relativistic outflows \cite{bm43},
which is known as funnel-wall jets. The base of the jet is enveloped
by the jet sheath that generates substantial thermal synchrotron
emission. As a result, the radiation observed in BH images may arise
from both the accretion flow and the jet. In recent years,
considerable attention has been devoted toward a range of accretion
disk models, including spherical accretion models
\cite{bm44,Zeng:2020dco,GB28,gkref26,sd38}, geometrically thin
accretion disks \cite{bm45,sd34,sd35,bm46,sdn1,israrthin}, and
Einstein rings in holography \cite{sd21,sd22,sd23,israr1}.

As the central engines of many low-luminosity active galactic
nuclei, supermassive BHs are often embedded in magnetized,
advection-dominated accretion matter \cite{bm47,bm48}. In these
environments, the plasma is characterized by strong radial inflow
and weak Coulomb coupling between ions and electrons \cite{bm49}.
General-relativistic magnetohydrodynamic (GRMHD) simulations,
although computationally intensive, provide the most self-consistent
mechanism currently available for studying these accretion flows
\cite{bm50}, and can be investigated in modified theories
\cite{bm51,bm52}. In this scenario, various studies considers the
radiatively inefficient accretion flow (RIAF) model with relatively
low radiation efficiency, in which the vertically averaged electron
number density and temperature typically follow a power-law
distribution with radius \cite{GB35,GB36,GB37,GB38}.

On the other hand, semi-analytic accretion models continue to play a
valuable complementary role. They allow for efficient exploration of
parameter space and enable controlled investigations that isolate
specific physical processes \cite{bm53,bm54}. However, many such
models are built on simplifying assumptions, such as self-similar
radial profiles, geometrically thin disks, or nearly circular orbits
that may break down in the strongly relativistic regions close to
the event horizon. In the immediate vicinity of the event horizon,
gravity dominates the dynamics, making the ballistic approximation
treating the plasma motion as timelike geodesics, provides a useful
and analytically manageable description of the flow
\cite{bm55,bm56}. Based on this mechanism, Hou et al. \cite{bm57}
developed a self-consistent model (known as ballistic approximation
accretion flow (BAAF) model) in Kerr spacetime, which provides
explicit expressions for thermodynamic quantities and magnetic field
configurations. Consequently, this model provides a physically
consistent mechanism for describing the structure and dynamics of
geometrically thick accretion flows, and enabling comprehensive
investigation of polarization characteristics in the near horizon
region.

The thermal synchrotron radiation radiated by electrons in BH
accretion flows having different polarization properties. The
polarization vector is perpendicular to both the magnetic field and
the direction of wave propagation, aligning with the electric field
vector. In the intense gravitational field, this vector is carried
along null geodesics through parallel transport and forming the
polarized image of the accretion flow on the observer's screen. As a
result, polarimetric measurements serve as a powerful probe of the
plasma behavior and magnetic field geometry in the vicinity of BHs
\cite{bm58}. Recent polarization images released by EHT have display
a prominent polarization structures within emission rings
\cite{bm59,bm60}. Using the approximate analytic ray-tracing
formalism \cite{bm61}, in 2021, EHT regenerated both the electric
vector position angle (EVPA) distribution and the relative
polarization intensity of M$87^{\ast}$ \cite{bm62}. In this
perspectives, the shadows and polarization images of various BH
models, including horizonless ultra compact objects, have been
extensively explored in the literature
\cite{young1,th36,young2,th37,th38,th39,zengnew1,zengnew2,GB45,GB46,israrGBthick,thickstar,horndskithick}.
Building upon this framework, we extend the analysis to
Kerr-Sen-like metric in Einstein-Bumblebee gravity and investigate
the associated consequences for BH imaging illuminated by a thick
accretion flow. For desired results, we consider RIAF and BAAF
accretion models. We examine how the LSB parameter $\ell$, the
charge parameter $Q$, and the observer inclination modulates key
flow properties, including the density profile, thermal structure,
and magnetic-field configuration. We further examine the
polarization properties and explore how these parameters influence
the polarization structure near the event horizon.

The remaining parts of this paper are structured as follows. In
sect. {\bf II}, we briefly define the Kerr-Sen-like metric coming
from Einstein-Bumblebee gravity and the corresponding geodesics
equations including the definition of photon spheres. In sect. {\bf
III}, we describe the fundamental mechanism responsible for
synchrotron radiation and the covariant form of the radiative
transfer equation, including both the isotropic and anisotropic
radiation models. Section {\bf IV} reflects the background of the
accretion flow models consists of RIAF and BAAF frameworks. We
outline the theoretical mechanism for polarization imaging within
the BAAF model in sect. {\bf V}. The last section offers our
concluding remarks. Throughout this paper, we adopt the natural
units with $G=M=c=1$ without loss of generality.

\section{Kerr-Sen-like Black Hole and Null Geodesics}
We describe the Kerr-Sen-like BH metric, whose axisymmetric rotating
BH solution in Boyer-Lindquist coordinates can be written as
\cite{bm27,bm30,bm37}
\begin{eqnarray}
ds^2&=&-\bigg(1-\frac{2r}{\Sigma}\bigg)dt^2+(1+\ell)\frac{\Sigma}{\Delta}dr^2+\Sigma
d\theta^2+\frac{A\sin^{2}\theta}{\Sigma}d\phi^2-\frac{4\tilde{A}r\sin^{2}\theta}{\Sigma}dt
d\phi,\label{bmg1}
\end{eqnarray}
where\footnote{A more widely used notation is
$\Delta=\frac{r(r+Q)-2r}{1+\ell}+a^{2}$.}
\begin{equation}
\Sigma=r(r+Q)+\tilde{A}^{2}\cos^{2}\theta, \quad
\Delta=r(r+Q)-2r+\tilde{A}^{2},\quad
A=[r(r+Q)+\tilde{A}^{2}]^{2}-\tilde{A}^{2}\Delta\sin^{2}\theta,
\end{equation}
with $\tilde{A}=a\sqrt{1+\ell}$. Moreover, the parameter $\ell$
represents the LSB, and $Q$ represent the charge parameter. In the
framework of Einstein-Bumblebee gravity, $Q$ can be regarded as the
vacuum expectation value of the Bumblebee field, and $\ell\propto
\varrho Q^{2}/a$, where $\varrho$ is the coupling constant between
the gravitational field and the Bumblebee field and $a$ is the spin
parameter \cite{bm27,bm30,bm37}. The event horizons are determined
by $\Delta/\Sigma=0$, which yields
\begin{equation}
r_{\pm}=\frac{2-Q\pm\sqrt{4-4\tilde{A}^{2}-4Q+Q^{2}}}{2},
\end{equation}
where $r_+$ and $r_-$ correspond to the event horizon and the Cauchy
horizon, respectively. Moreover, the spinning Kerr-Sen-like BH
(\ref{bmg1}) exhibits time-translational and axial rotational
symmetries, which imply the presence of two corresponding Killing
vector fields. To facilitate the study of BH shadow imaging, we
employ the zero angular momentum observer (ZAMO) frame, defined as a
stationary observer with vanishing angular momentum at infinity.
However, due to the effect of frame dragging, the ZAMO still has a
position dependent angular velocity
$\omega=\frac{d\phi}{dt}=-\frac{g_{t\phi}}{g_{\phi\phi}}$. As the
observer reaches the BH, $\omega$ increases and reaches its maximum
at the event horizon $\Omega=\omega|_{r=r_{+}}$, in which $\Omega$
interpret the angular velocity of the BH. At this position, the
observer's angular velocity $\omega$ coincides with $\Omega$
indicating a state of corotation. Now we evaluate the motion of
light particles in the vicinity of spinning Kerr-Sen-like BH. Since
the photon follows the null geodesics in a given BH spacetime,
hence, for spacetime (\ref{bmg1}), the geodesic motion is governed
by the Hamilton-Jacobi equation, which is given by \cite{bm63}
\begin{eqnarray}\label{hd3}
\frac{\partial \tilde{\mathcal{I}}}{\partial
\varpi}=-\frac{1}{2}g^{\mu\nu}\frac{\partial
\tilde{\mathcal{I}}}{\partial x^{\mu}}\frac{\partial
\tilde{\mathcal{I}}}{\partial x^{\nu}},
\end{eqnarray}
here $\tilde{\mathcal{I}}$ indicates the Jacobi action and $\varpi$
is the affine parameter of the trajectory curves. The Jacobi action
$\tilde{\mathcal{I}}$ of the photon can be separated into the
following form
\begin{equation}\label{s4}
\tilde{\mathcal{I}}=\frac{1}{2}\varrho^2\varpi-E
t+L\phi+J_r(r)+J_{\theta}(\theta).
\end{equation}
For photons $\varrho=0$. The quantities $E=-p_{t}$ and $L=p_{\phi}$
represents the conserved energy and conserved angular momentum of
the photon in the direction of rotation axis, respectively. The
components $J_r(r)$ and $J_{\theta}(\theta)$ are arbitrary
functions. In this regard, the geodesic equations can be written in
first-order form
\begin{eqnarray}\nonumber
\Sigma^{2}\frac{dt}{d\varpi}&=&a(L-aE\sin^{2}\theta)+\frac{r^{2}+a^{2}}{\Delta}(E(r^{2}+a^{2})-a
L),\\\nonumber
\Sigma^{2}\frac{dr}{d\varpi}&=&\pm\sqrt{\tilde{R}(r)},\\\nonumber
\Sigma^{2}\frac{d\theta}{d\varpi}&=&\pm\sqrt{\Theta(\theta)},\\\label{s5}
\Sigma^{2}\frac{d\phi}{d\varpi}&=&(L\csc^{2}\theta-aE)+\frac{a}{\Delta}(E(r^{2}+a^{2})-aL).
\end{eqnarray}
Here
\begin{eqnarray}\nonumber
\tilde{R}(r)&=&(E(r^{2}+a^{2})-aL)^{2}-\Delta(\tilde{Q}+(L-aE)^{2}),\\\label{s6}
\Theta(\theta)&=&\tilde{Q}+\big(a^{2}E^{2}-L^{2}\csc^{2}\theta\big)\cos^{2}\theta,
\end{eqnarray}
are the radial and angular potentials, respectively, while
$\tilde{Q}$ is the Carter constant. On the equatorial plane, where
$\theta=\pi/2$, the radial equation can be rewritten as
$\dot{r}^{2}+V_{\text{eff}}=0$, where $V_{\text{eff}}$ is the
effective potential, which is defined as
\begin{equation}\label{vb1}
V_{\text{eff}}=-\frac{\tilde{R}(r)}{r^{4}}.
\end{equation}
For convenience, we define the dimensionless impact parameters of
photons
\begin{eqnarray}\label{s6}
\xi=\frac{L}{E}, \quad \quad \eta = \frac{\tilde{Q}}{E^2}.
\end{eqnarray}
The photon sphere radius $r_{p}$ can be evaluated from the radial
potential and its derivative as \cite{bm63}
\begin{eqnarray}\label{sn6}
\tilde{R}(r)|_{r=r_{p}}=0, \quad \quad
\partial_{r}\tilde{R}(r)|_{r=r_{p}}=0.
\end{eqnarray}
Considering a photon captured by the BH, those residing on stable
orbits remain bound for infinitely, whereas photons on unstable
orbits may eventually escape after circling the BH for some time.
The conditions for such unstable orbits are given by
\begin{equation}\label{vbn2}
\partial^{2}_{r}V_{\text{eff}}|_{r=r_{p}}<0.
\end{equation}

\section{Synchrotron Radiation and Radiative Transfer}\label{sec3}

We consider the presence of a magnetic field in the vicinity of the
BH, the explicit configuration of which will be specified in
subsequent sections. Within this field, both thermal and nonthermal
electrons in the accreting plasma emit synchrotron radiation as they
are accelerated by the Lorentz force. Our investigation is primarily
concerned with synchrotron emission arising from highly relativistic
electrons. In this section, we present an overview of the
fundamental mechanism responsible for synchrotron radiation in the
surrounding plasma and describe the propagation of the emitted
photons from their source to the observer's image plane. All
physical quantities are expressed in CGS units. For unpolarized
environment, the covariant form of the radiative transfer equation
is \cite{Gold:2020iql}
\begin{align}
    \label{eq:radiative_transfer}
    \frac{d}{d\varpi} I = J - \lambda I.
\end{align}
where, $I$, $J$, and $\lambda$ are Lorentz invariant quantities. For
an arbitrary observer, let $\nu$ represents the photon frequency
measured in the observer's frame. The relations between these
invariants are defined as
\begin{align}
    \label{eq:invariant_relations}
    I = \frac{I_\nu}{\nu^3}, \quad J = \frac{j_\nu}{\nu^2}, \quad \lambda = \nu \lambda_\nu.
\end{align}
in which $I_\nu$ is the specific intensity, $j_\nu$ is the
emissivity, and $\lambda_\nu$ is the absorption coefficient. The
solution of Eq.~\eqref{eq:radiative_transfer}   is
\begin{align}
    I(\varpi) = I(\varpi_0) + \int_{\varpi_0}^{\varpi} d\varpi' \, J(\varpi') \exp\left( -\int_{\varpi'}^{\varpi} d\varpi'' \, \lambda(\varpi'') \right).
\end{align}

To ensure consistency with the unit system adopted in
Eq.~\eqref{eq:invariant_relations}, we recast the equation in the
Centimeter-Gram-Second (CGS) system of units. For this purpose, we
rescale the affine parameter $\varpi$ appearing in
Eq.~\eqref{eq:radiative_transfer} according to the transformation
$\frac{d}{d\varpi} \rightarrow \frac{1}{C}\frac{d}{d\varpi}$, where
$C = \frac{r_g}{\nu_0}$. Here, $r_g = \frac{GM}{c^2}$ denotes the
gravitational radius of the black hole, and $\nu_0$ represents the
photon frequency as measured by an observer at infinity. With this
rescaling, the equation takes the following form

\begin{align}
    \frac{1}{C} \frac{d}{d\varpi} I = J - \lambda I,
\end{align}
with the corresponding solution
\begin{align}\label{eq:specific_intensity}
I_\nu = g^3 I_{\nu_0} + r_g \int_{\varpi_0}^{\varpi} d\varpi' \, g^2
j_\nu(\varpi') \exp\left( -r_g \int_{\varpi'}^{\varpi} d\varpi'' \,
\frac{\lambda_\nu(\varpi'')}{g} \right).
\end{align}
Here, $g = \nu_0 / \nu$ is the redshift factor, and $\nu$
corresponds to the photon frequency measured in the local static
frame. For explicit expression of $g$, we define the fluid four
velocity be $u^{\alpha}$ and the photon four-momentum be
$k_{\alpha}$ with $k_t = -1$, then
\begin{align}
g = \frac{k_\alpha (\partial_t)^\alpha}{k_\alpha u^\alpha} =
\frac{k_t}{k_\alpha u^\alpha} = - \frac{1}{k_\alpha u^\alpha}.
\end{align}
From the above relations, it is evident that both the emissivity and
the absorption coefficient must be specified in order to accurately
evaluate the observed intensity. We notice that the radiative
coefficients $j_{\nu}$ and $\lambda_\nu$ as defined in
Eq.~\eqref{eq:invariant_relations} closely depend on the specific
emission mechanism, with different physical processes leading to
distinct functional forms of $j_{\nu}$ and $\lambda_\nu$. In present
analysis, we focus on synchrotron radiation produced by electrons in
the ultra-relativistic regime, expressed in CGS units. Throughout
this section, $c$ denotes the speed of light, $h$ is the Planck
constant, $e$ is the elementary charge, and $k_B$ represent the
Boltzmann constant, while the local magnetic field is represented by
$b^\alpha$.

In a plasma system, synchrotron radiation is mainly participated by
electrons, and its emissivity $j_{\nu}$ plays a major role in thick
disk imaging, given explicitly by
\begin{align}
    \label{eq:emissivity}
    j_\nu = \frac{\sqrt{3} e^3 B \sin\theta_B}{4\pi m_e c^2} \int_0^\infty d\tau \, N(\tau) F\left( \frac{\nu}{\nu_s} \right).
\end{align}
Here, $\tau = \frac{1}{\sqrt{1 - \beta^2}}$ denotes the Lorentz
factor of the charged particle, $N(\tau)$ corresponds to the
electron configuration function, and $F(x)$ is expressed as
\begin{align}\label{eq:F_function}
F(x) = x \int_x^\infty dy\, K_{5/3}(y),
\end{align}
where $K_n(x)$ is the modified Bessel function of the second kind of
order $n$. The angle $\theta_B$ between the spatial projection of
the photon four-momentum $e^{\alpha}_{(k)}$ and the magnetic field
direction $e^{\alpha}_{(b)}$ is
\begin{align}
    \theta_B = \arccos\left(e_{(b)}^\alpha \cdot e_{(k)}^\alpha\right) = \arccos\left[\frac{g}{B}(b_\alpha k^\alpha)\right],
\end{align}

with
\begin{align}\label{eq:four_vectors}
e^\alpha_{(k)} = -\left( \frac{k^\alpha}{u^{\nu} k_\nu} + u^\alpha
\right), \quad e^\alpha_{(b)} = \frac{b^\alpha}{B}.
\end{align}
Here, $B = \sqrt{b_\alpha b^\alpha}$ is the magnitude of the local
magnetic field. In Eq.~\eqref{eq:emissivity}, the characteristic
frequency $\nu_s$ is
\begin{align}
    \nu_s = \frac{3 e B \tau^2 \sin\theta_B}{4 \pi m_e c}.
\end{align}
Different electron distributions give rise to distinct emissivities.
For a thermal electron population, the distribution function
$N(\tau)$ takes the following form

\begin{align}
    \label{eq:thermal_distribution}
    N(\tau) = \frac{n_e \beta \tau^2}{\theta_e K_2(\theta_e^{-1})} e^{-\tau / \theta_e},
\end{align}
where $n_e$ is the electron number density, $\theta_e = k_B T_e /
m_e c^2$ is the dimensionless electron temperature, and $T_e$
denotes the thermodynamic temperature of electrons. In the extreme
relativistic limit, $\beta \approx 1$ and $\theta_e \gg 1$, the
asymptotic form $K_2(\frac{1}{\theta_e}) \approx 2 \theta_e^2$
holds. Defining $z = \tau / \theta_e$, Eq.~\eqref{eq:emissivity}
becomes as
\begin{align}
    j_\nu = \frac{\sqrt{3} n_e e^3 B \sin\theta_B}{8 \pi m_e c^2} \int_0^\infty z^2 e^{-z} F\left( \frac{\nu}{\nu_s} \right) dz.
\end{align}
Defining $x = (\nu / \nu_s) z^2$, the emissivity can be defined as
\begin{align}
    \label{eq:anisotropic_emissivity}
    j_\nu = \frac{\sqrt{3} n_e e^2 \nu}{6 c \theta_e^2} \mathcal{I}(x), \quad x = \frac{\nu}{\nu_c}, \quad \nu_c = \frac{3 e B \theta_e^2 \sin\theta_B}{4 \pi m_e c},
\end{align}
where $\mathcal{I}(x)$ is the dimensionless function, which is
defined as
\begin{align}
    \mathcal{I}(x) = \frac{1}{x} \int_0^\infty z^2 e^{-z} F\left( \frac{x}{z^2} \right) dz.
\end{align}
Since this function does not admit a closed-form expression in terms
of elementary functions, it is approximated using an appropriate
fitting formula. In this manuscript, we consider two radiation
models: isotropic and anisotropic radiation. We first discuss the
isotropic radiation model. For isotropic radiation, only the
magnitude of the magnetic field is considered, ignoring its
direction. The angle averaged synchrotron emissivity is defined by
\begin{align}
    \label{eq:angle_averaged_emissivity}
    \overline{j}_{\nu} = \frac{1}{2} \int_{0}^{\pi} j_{\nu} \sin\theta_B \, d\theta_B.
\end{align}
Its corresponding fitting formula, as provided in
Ref.~\cite{Mahadevan:1996cc}, reads
\begin{align}
    \overline{j}_{\nu} = \frac{n e^{2} \nu}{2\sqrt{3} c \theta_{e}^{2}} \mathbb{M}(x), \quad x = \frac{\nu}{\nu_c}, \quad \nu_c = \frac{3 e B \theta_e^2}{4\pi m_e c},
\end{align}
where the dimensionless function $\mathbb{M}(x)$ is given by
\begin{align}
    \mathbb{M}(x) = \frac{4.0505}{x^{1/6}} \left(1 + \frac{0.4}{x^{1/4}} + \frac{0.5316}{x^{1/2}}\right) \exp\left(-1.8899 x^{1/3}\right).
\end{align}
Next, we consider an anisotropic radiation model. In this mechanism,
the magnetic field is assumed to consist of a combination of
toroidal and poloidal components, such that the magnetic four-vector
can be expressed as follows
\begin{align}
    b^\alpha \sim (l, 0, 0, 1),
\end{align}
with
\begin{align}
    l = -\frac{u_\phi}{u_t}, \quad u_\nu = g_{\alpha\nu} u^\alpha = (u_t, u_r, u_\theta, u_\phi).
\end{align}
The magnetic field is perpendicular to the fluid four velocity,
satisfying $u^\alpha b_\alpha = 0$. The emissivity for the
anisotropic radiation model is given by
Eq.~\eqref{eq:anisotropic_emissivity}, i.e.,
\begin{align*}
j_\nu = \frac{n_e e^2 \nu}{2\sqrt{3} c \theta_e^2} \mathcal{I}(x),
\quad x = \frac{\nu}{\nu_c}, \quad \nu_c = \frac{3 e B \theta_e^2
\sin\theta_B}{4 \pi m_e c},
\end{align*}
where the dimensionless function $\mathcal{I}(x)$ is given in
Ref.~\cite{2011Numerical} as
\begin{align}
    \mathcal{I}(x) = 2.5651 \left( 1 + 1.92 x^{-1/3} + 0.9977 x^{-2/3} \right) \exp\left( -1.8899 x^{1/3} \right).
\end{align}
For a thermal electron distribution, the absorption process
satisfies Kirchhoff's law, so that the absorption coefficient
$\lambda_\nu$ obeys
\begin{align} \label{eq:blackbody}
 \lambda_\nu = \frac{j_\nu}{\mathcal{B}_\nu}, \quad \mathcal{B}_\nu = \frac{2 h \nu^3}{c^2} \frac{1}{\exp\left( \frac{h \nu}{k_B T_e} \right) -
 1},
\end{align}
where, $\mathcal{B}_\nu$ is the Planck black-body function. For
numerical simulations, we express the following constants, as
\begin{align}
    C_{1} = \frac{\sqrt{3} e^{2} n_{h} \nu_{h}}{6 \theta_{h}^{2} c}, \quad
    C_{2} = \frac{4 \pi c m_{e} \nu_{h}}{3 e B_{h} \theta_{h}^{2}}, \quad
    C_{3} = \frac{h \nu_{h}}{m_{e} \theta_{h} c^{2}}, \quad
    C_{4} = \frac{2 h \nu_{h}^{3}}{c^{2}}, \quad
    C_{5} = \sqrt{c^{2} n_{h} m_{p}},
\end{align}
in which, $n_h$, $\theta_h$, $\nu_h$, and $B_h$ corresponds to the
values of the electron number density, the dimensionless electron
temperature, the photon frequency, and the local magnetic field
strength at the event horizon, respectively. Particularly, we set
$\nu_h = 10^9~\mathrm{Hz} = 1~\mathrm{GHz}$ and $B_h = 1$. Based on
this parameterization, the emissivity and the black-body function
can be defined as
\begin{align}\label{eq:parameterized}
j_\nu = \frac{C_1 \hat{n}_e \hat{\nu} I(x)}{\hat{\theta}_e^2},
\qquad x = \frac{C_2 \hat{\nu}}{\hat{B} \hat{\theta}_e^2
\sin\theta_B}, \qquad \mathcal{B}_\nu = \frac{C_4
\hat{\nu}^3}{\exp\left( \frac{C_3 \hat{\nu}}{\hat{\theta}_e} \right)
- 1},
\end{align}
here $\hat{\nu} = \nu / \nu_h$, $\hat{n}_e = n_e / n_h$, $\hat{B} =
B / B_h$, and $\hat{\theta}_e = \theta_e / \theta_h$. Using
Eqs.~\eqref{eq:blackbody} and \eqref{eq:parameterized}, one can
compute the intensity in Eq.~\eqref{eq:specific_intensity}. However,
the electron number density and temperature appearing in these
expressions remain unspecified. In the following section, we outline
how these quantities are determined within the context of different
accretion flow models.
\section{Disk Models}\label{sec4}
For unpolarized imaging, we examine two geometrically thick and
optically thin accretion flow models, well-known as RIAF
model~\cite{Broderick:2010kx} and the BAAF
model~\cite{bm57,Zhang:2024lsf}. The RIAF model is broadly
consistent with GRMHD simulations and has demonstrated considerable
success in reproducing the overall morphology of M87*~\cite{bm1}.
However, its applicability to polarization studies remains limited,
as it neglects key physical ingredients such as outflows,
non-thermal particles, and the full dynamical structure captured by
GRMHD frameworks. To overcome these limitations, we adopt the BAAF
model, which assumes that fluid acceleration in the vicinity of the
event horizon is predominantly governed by gravitational effects.
This model provides explicit prescriptions for the thermodynamic
variables and magnetic field configuration, enabling a more
realistic and self-consistent description of the morphology and
dynamics of geometrically thick accretion flows in the near-horizon
region of the BH.
\subsection{Density and Temperature}
We impose a cylindrical coordinate system, where the cylindrical
radius is given by $R = r \sin\theta$ and the height above the
equatorial plane ($\theta = \pi/2$) is $z = r \cos\theta$. Closely
followed by~\cite{Broderick:2010kx}, in RIAF model, the density and
temperature profiles can be expressed as
\begin{align}
    n_e = n_h \left( \frac{r}{r_+} \right)^2 \exp\left( -\frac{z^2}{2 R^2} \right), \quad
    T_e = T_h \left( \frac{r}{r_+} \right),
\end{align}
where $n_h$ and $T_h$ are the electron number density and
temperature at the outer horizon, respectively. The magnetic field
magnitude is defined via the cold magnetization parameter $\eta =
\frac{b^2}{\rho} = \frac{b^2}{n_e m_p c^2}$ as $b = \sqrt{\eta
\rho}$, where $\rho = n_e m_p c^2$ is the fluid mass density, and
$\eta \sim 0.1$~\cite{Pu:2016qak}. Here, we consider both the
isotropic radiation~\eqref{eq:angle_averaged_emissivity} and
anisotropic radiation~\eqref{eq:anisotropic_emissivity} mechanism.
After specifying the fundamental properties of the accretion flow
motion, we proceed to describe its dynamics. In general, the fluid
motion may be modeled as free-fall, circular motion around the BH,
or a combination of both. In the present work, we primarily focus on
the free-fall scenario. Under the assumption that the fluid is
initially at rest at infinity, the corresponding four-velocity can
be written as:
\begin{eqnarray}
u_t = -g^{tt}, \quad u^r = -\sqrt{-(1 + g^{tt})g^{rr}}.
\end{eqnarray}
The four-velocity must be a timelike vector throughout the entire
space-time, requiring $g^{tt} \leq -1$. After establishing the
fundamental properties and dynamics of the accreting matter, we
proceed to describe the behavior of electrons through the radiative
transfer equation. Further, we employ a ray-tracing method combined
with a ZAMO frame and celestial coordinates to establish the mapping
between pixel coordinates on the projection screen and celestial
coordinates. This provides a solid computational foundation for BH
imaging studies. For comprehensive review, one can see Refs.
\cite{young1,sd35}.

\begin{figure}[H]
\centering \subfigure[$Q=0.1,
\ell=-0.5$]{\includegraphics[scale=0.45]{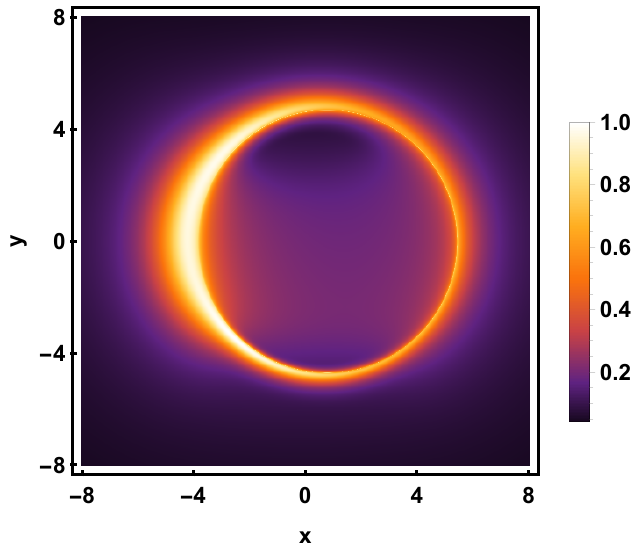}}
\subfigure[$Q=0.1,
\ell=0.1$]{\includegraphics[scale=0.45]{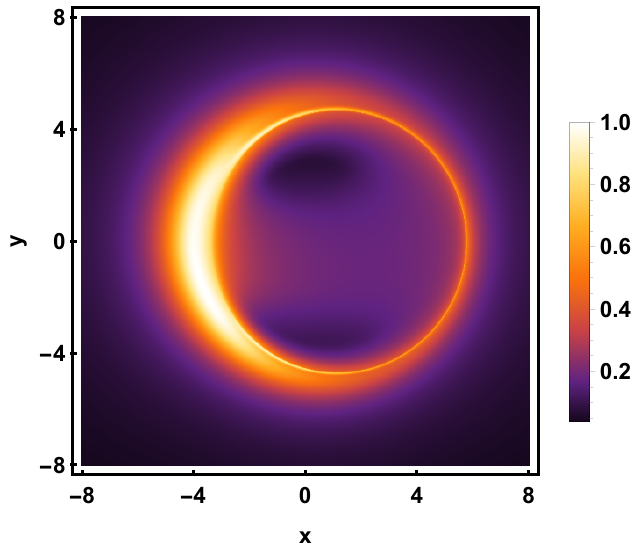}}
\subfigure[$Q=0.1,
\ell=0.5$]{\includegraphics[scale=0.45]{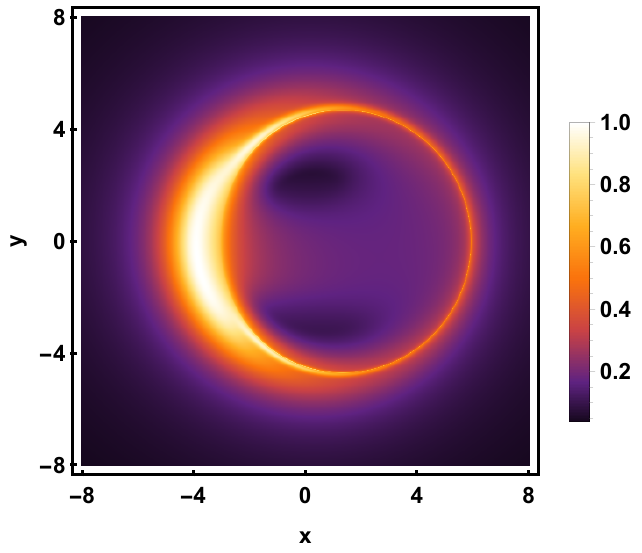}}
\subfigure[$Q=0.3,
\ell=-0.5$]{\includegraphics[scale=0.45]{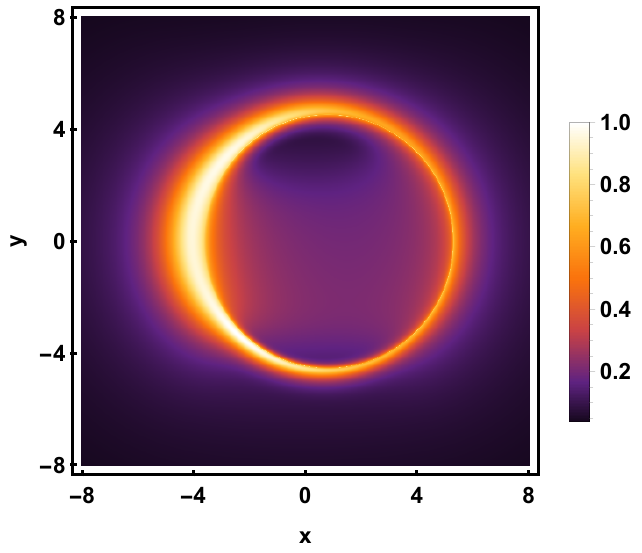}}
\subfigure[$Q=0.3,
\ell=0.1$]{\includegraphics[scale=0.45]{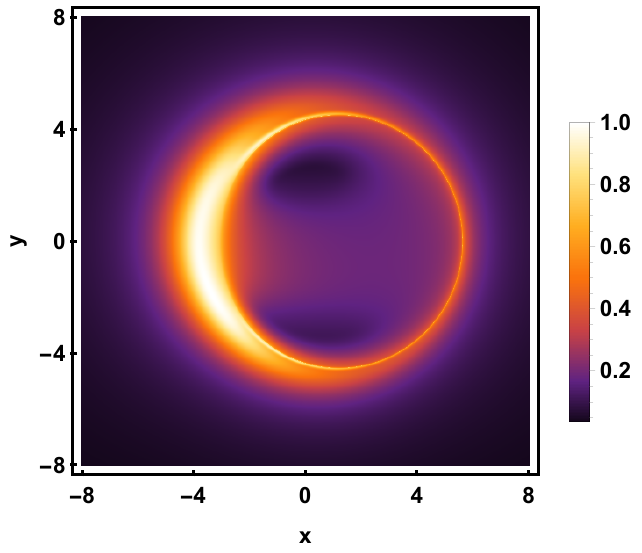}}
\subfigure[$Q=0.3,
\ell=0.5$]{\includegraphics[scale=0.45]{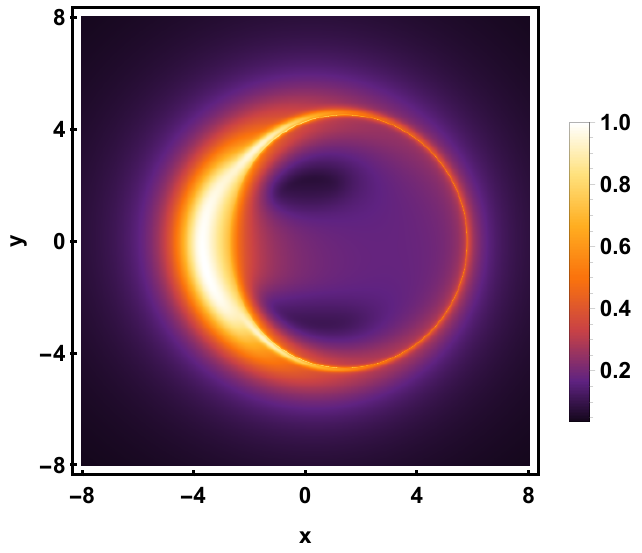}}
\subfigure[$Q=0.5,
\ell=-0.5$]{\includegraphics[scale=0.45]{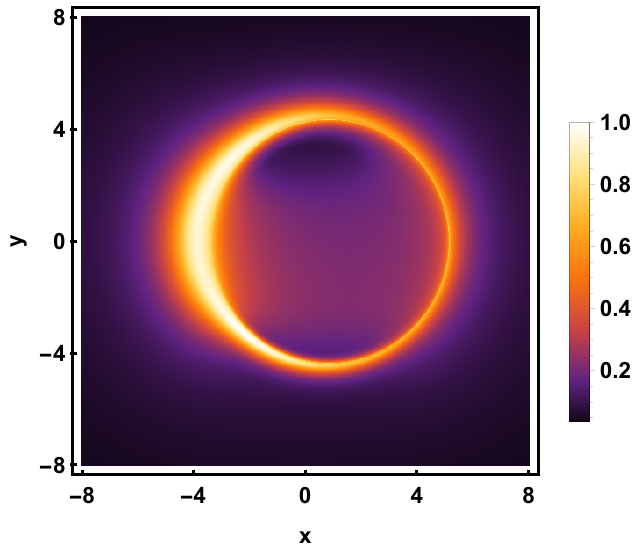}}
\subfigure[$Q=0.5,
\ell=0.1$]{\includegraphics[scale=0.45]{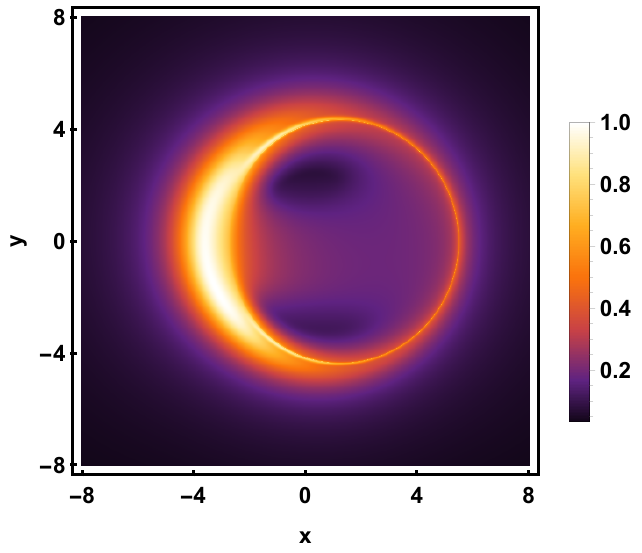}}
\subfigure[$Q=0.5,
\ell=0.5$]{\includegraphics[scale=0.45]{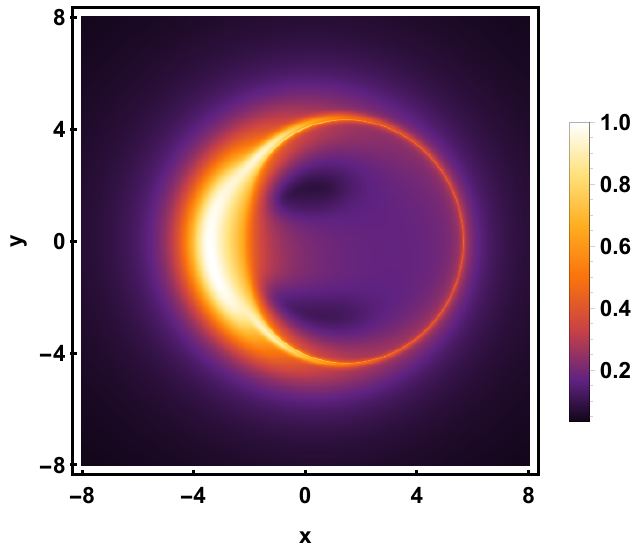}} \caption{Imaging
results of the thick accretion disk for isotropic radiation under
the RIAF model, and the accretion flow follows the infalling motion.
The observation is performed at a frequency of $230\mathrm{GHz}$
with an inclination angle of $\theta_o=75^\circ$, and a spin
parameter $a=0.6$.}\label{figtong1}
\end{figure}

\begin{figure}[H]\centering \subfigure[Horizontal
direction]{\includegraphics[scale=0.8]{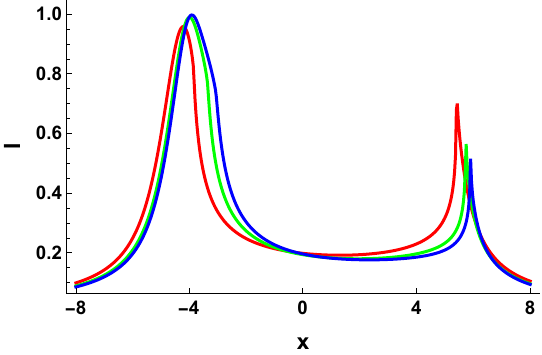}}
\subfigure[Vertical
direction]{\includegraphics[scale=0.8]{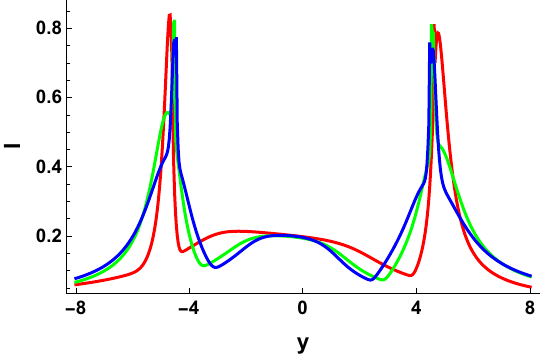}} \caption{Intensity
distribution for isotropic radiation under the RIAF model, and the
accretion flow follows the infalling motion. The red, green and blue
curves corresponds to $\ell=-0.5,~0.1$ and $0.5$, respectively with
a fixed $a=0.6,~Q=0.1,~\theta_o=75^\circ$.} \label{figtong_curve1}
\end{figure}

\begin{figure}[H]\centering \subfigure[Horizontal
direction]{\includegraphics[scale=0.8]{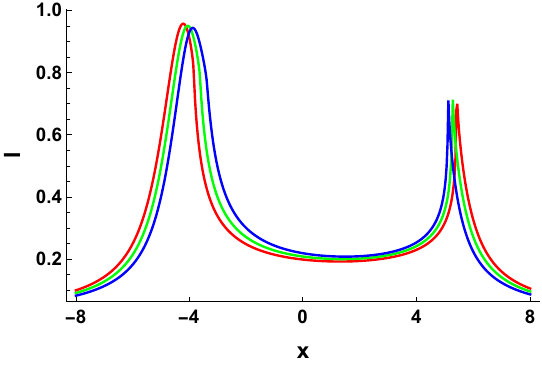}} \subfigure[Vertical
direction]{\includegraphics[scale=0.8]{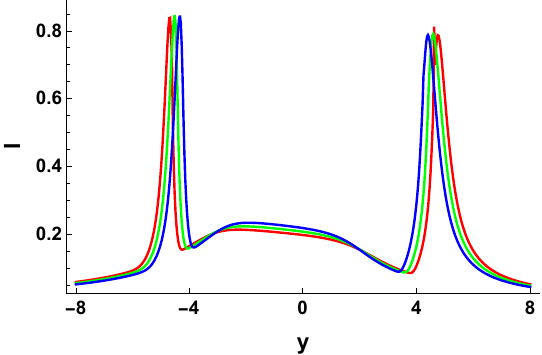}} \caption{Intensity
distribution for isotropic radiation under the RIAF model, and the
accretion flow follows the infalling motion. The red, green and blue
curves corresponds to $Q=0.1,~0.3$ and $0.5$, respectively with a
fixed $a=0.6,~\ell=-0.5,~\theta_o=75^\circ$.} \label{figtong_curve2}
\end{figure}
In Fig. \textbf{\ref{figtong1}}, we interpret the shadow images of
Kerr-Sen-like BH with isotropic radiation under RIAF model. The
observation is performed at a frequency of $230\mathrm{GHZ}$ with an
inclination angle of $\theta_o=75^\circ$, and a spin parameter
$a=0.6$. The accretion flow follows the infalling motion inspired by
ballistic approximation. From left to right, the image plots for the
values of LSB parameter $\ell$ are $-0.5$, $0.1$, and $0.5$,
respectively. Whereas, from top to bottom, the image plots for the
values of charge parameter $Q$ are $0.1$, $0.3$, and $0.5$,
respectively. Analysis of Fig. \textbf{\ref{figtong1}} shows that
all images exhibit a bright ring structure, corresponding to
higher-order images, in which photons orbit the BH one or multiple
times before reaching the observer. This feature is a direct
manifestation of strong gravitational lensing. Beyond of this
ring-like structure, there are regions with nonzero intensity
corresponding to the primary image, where photons travel directly
from the accretion flow to the observer without orbiting the BH.
Moreover, regardless of parameter alterations, there are regions of
reduced intensity inside the higher-order images. This region
originates from the event horizon of the BH. For geometrically thin
accretion disk, the accreting matter is confined to the equatorial
plane, so the event horizon appears as a dark region, which is
well-known as ``inner shadow'', and may be captured by EHT
\cite{Chael:2021rjo}. On the contrary, for geometrically thick
accretion disks, radiation from outside the equatorial plane may
partially obscure this region, making it less distinguishable.
Compared to thin disks, thick disks are more physically realistic,
which explains why direct imaging of BH event horizons remains
challenging.

From the first row of Fig. \textbf{\ref{figtong1}}, when $Q=0.1$
(see first row), the higher-order images of BH exhibits a
``D''-shape in the left side of the screen, which is significantly
enhanced intensity with increasing $\ell$. As $Q$ increases to $0.3$
(see second row), the intensity of higher-order images are more
pronounced in the left side of the screen, which is increases with
increasing $\ell$. When $Q$ further increases to $0.5$ (see third
row), a crescent-shaped bright region appears on the left side of
the image, which is significantly enhanced with the aid of $\ell$.
Interestingly, when $\ell=0.1$, two dark regions appear inside the
higher-order image, with the upper region slightly darker than the
lower one, which is more obvious when $\ell=0.5$. This phenomenon
arises from gravitational lensing effects. In summary, increasing
$Q$ slightly decreases both the size and brightness of the
higher-order image, while increasing $\ell$ alters the shape of the
higher-order image and obscure the horizon's outline. Additionally,
the image is nearly symmetric in the vertical direction, although
the left-right intensity remains higher than the top-bottom
intensity. This vertical intensity dependence reflects the
equatorial symmetry of the thick disk: for observers near the
equatorial plane, high-latitude radiation partially fills the dark
regions, while for near-polar observers, photons reaching the
observer are relatively insufficient.

For a better understanding about Fig. \textbf{\ref{figtong1}}, we
have plotted Figs. \textbf{\ref{figtong_curve1}} and
\textbf{\ref{figtong_curve2}}, which exhibits the corresponding
intensity distributions along the $x$-axis and $y$-axis of the
observer's screen. All panels in Fig. \textbf{\ref{figtong1}}
exhibit a pronounced bright ring, corresponding to the peaks in
Figs. \textbf{\ref{figtong_curve1}} and
\textbf{\ref{figtong_curve2}}. Comparing the rows of Fig.
\textbf{\ref{figtong1}} with Fig. \textbf{\ref{figtong_curve1}}, we
observe that for a fixed value of $Q$, both the bright ring and
central region are slightly reduces with the augmentation of $\ell$.
Additionally, due to the frame-dragging effect induced by LSB
parameter $\ell$, the intensity on the left side of the image
becomes significantly enhanced with increasing $\ell$. On the
contrary, by comparing the columns of Fig. \textbf{\ref{figtong1}}
with those of Fig. \textbf{\ref{figtong_curve2}}, we observe that,
for a fixed value of $\ell$, both the bright ring and the central
dark region slightly shrink as $Q$ increases. Additionally, the
width of the ring decreases marginally.

\begin{figure}[H]
\centering \subfigure[$Q=0.1,
\ell=-0.5$]{\includegraphics[scale=0.45]{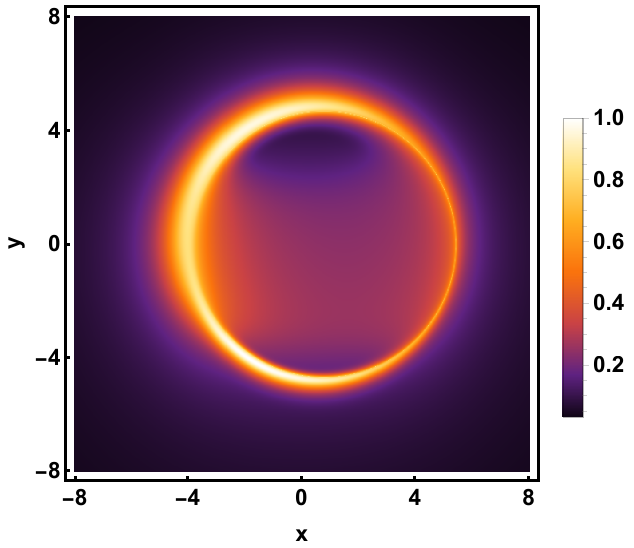}}
\subfigure[$Q=0.1,
\ell=0.1$]{\includegraphics[scale=0.45]{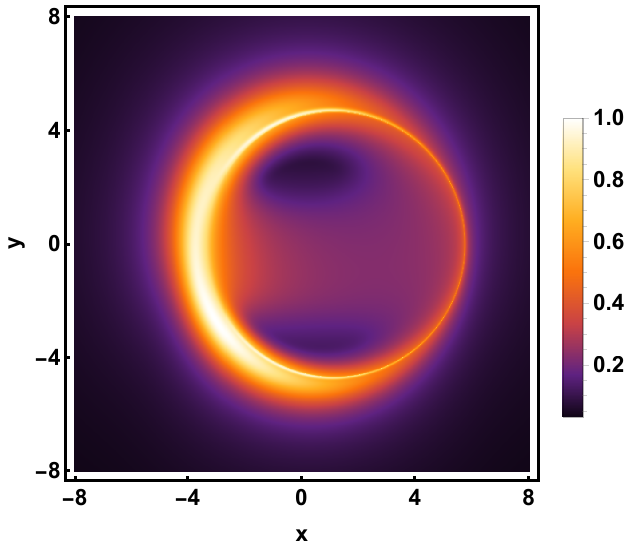}}
\subfigure[$Q=0.1,
\ell=0.5$]{\includegraphics[scale=0.45]{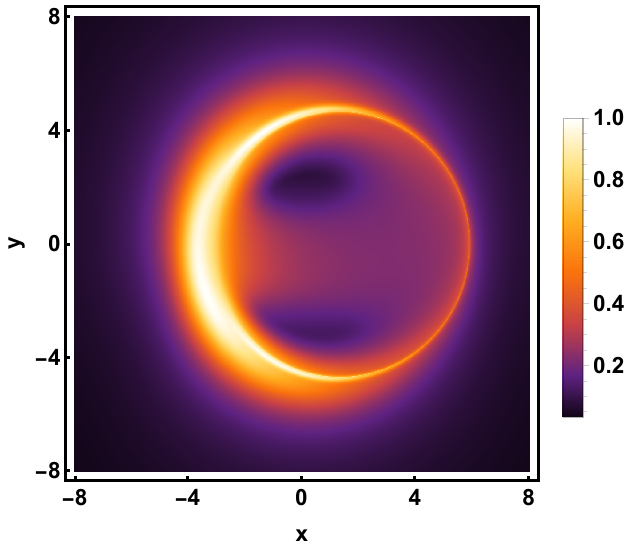}}
\subfigure[$Q=0.3,
\ell=-0.5$]{\includegraphics[scale=0.45]{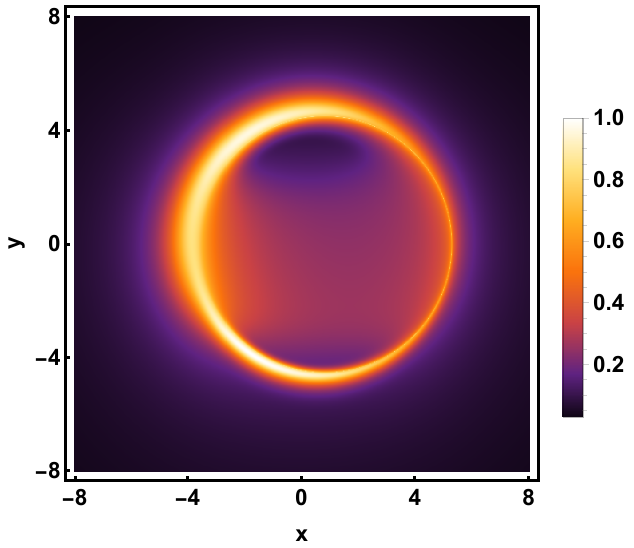}}
\subfigure[$Q=0.3,
\ell=0.1$]{\includegraphics[scale=0.45]{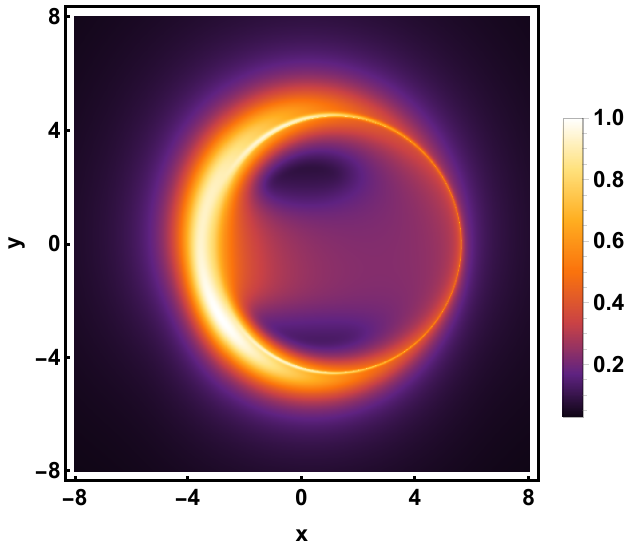}}
\subfigure[$Q=0.3,
\ell=0.5$]{\includegraphics[scale=0.45]{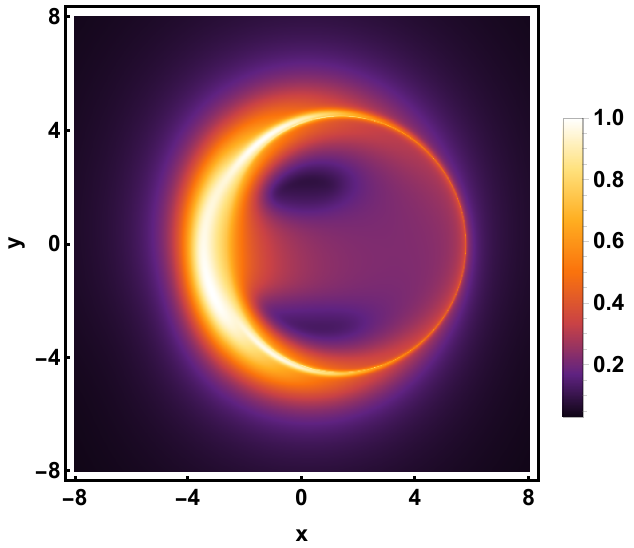}}
\subfigure[$Q=0.5,
\ell=-0.5$]{\includegraphics[scale=0.45]{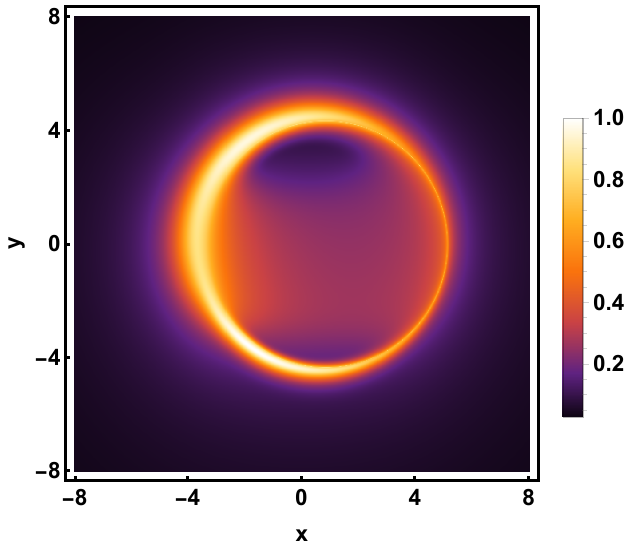}}
\subfigure[$Q=0.5,
\ell=0.1$]{\includegraphics[scale=0.45]{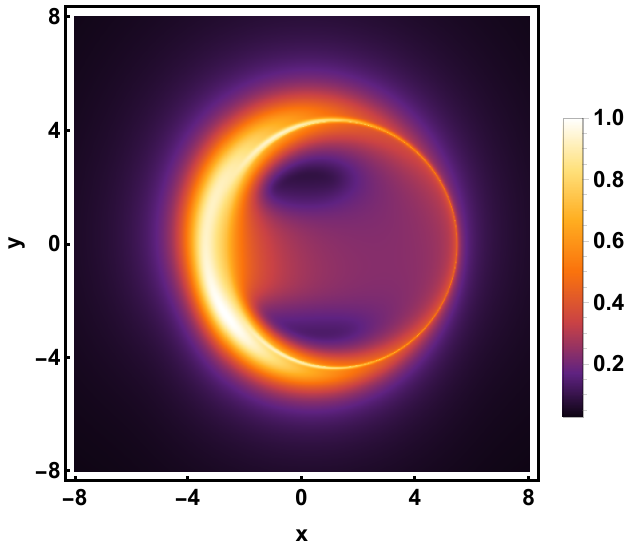}}
\subfigure[$Q=0.5,
\ell=0.5$]{\includegraphics[scale=0.45]{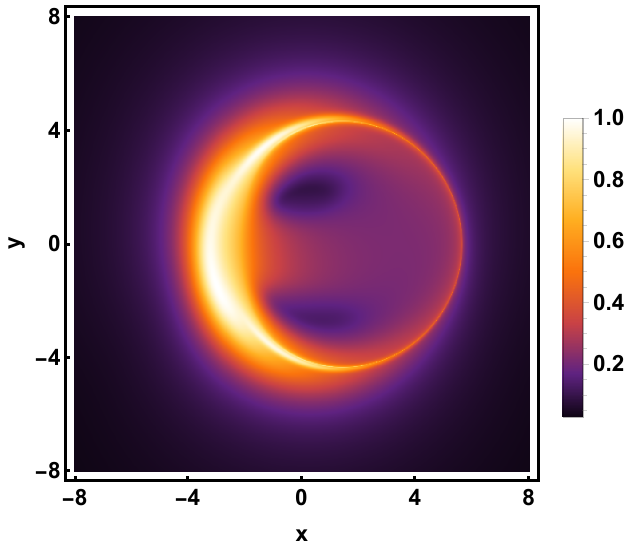}} \caption{Imaging
results of the thick accretion disk for anisotropic radiation under
the RIAF model, and the accretion flow follows the infalling motion.
The observation is performed at a frequency of $230\mathrm{GHz}$
with an inclination angle of $\theta_o=75^\circ$, and a spin
parameter $a=0.6$.}\label{figtong2}
\end{figure}

\begin{figure}[H]\centering \subfigure[Horizontal
direction]{\includegraphics[scale=0.8]{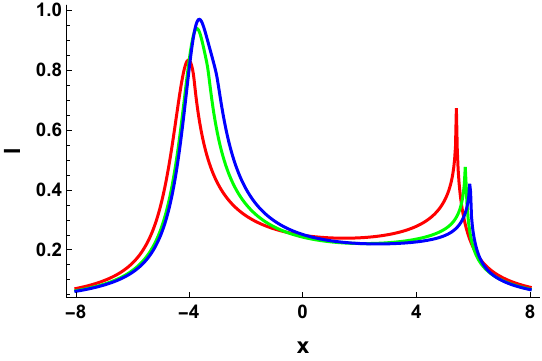}} \subfigure[Vertical
direction]{\includegraphics[scale=0.8]{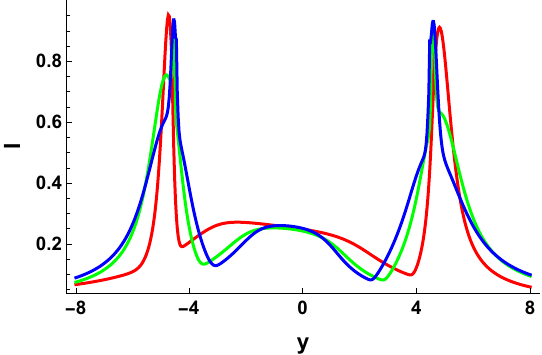}} \caption{Intensity
distribution for anisotropic radiation under the RIAF model, and the
accretion flow follows the infalling motion. The red, green and blue
curves corresponds to $\ell=-0.5,~0.1$ and $0.5$, respectively with
a fixed $a=0.6,~Q=0.1,~\theta_o=75^\circ$.} \label{figtong_curve3}
\end{figure}

\begin{figure}[H]\centering \subfigure[Horizontal
direction]{\includegraphics[scale=0.8]{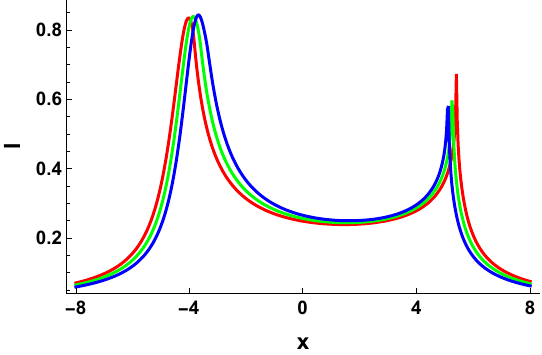}} \subfigure[Vertical
direction]{\includegraphics[scale=0.8]{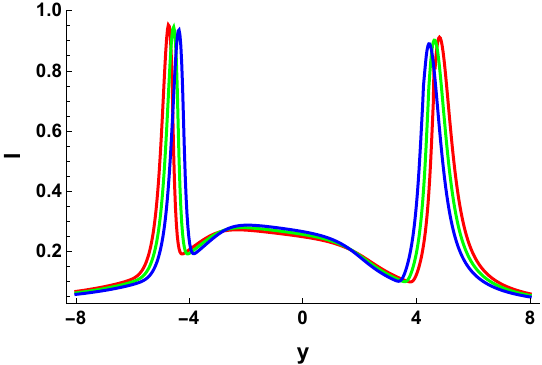}} \caption{Intensity
distribution for anisotropic radiation under the RIAF model, and the
accretion flow follows the infalling motion. The red, green and blue
curves corresponds to $Q=0.1,~0.3$ and $0.5$, respectively with a
fixed $a=0.6,~\ell=-0.5,~\theta_o=75^\circ$.} \label{figtong_curve4}
\end{figure}

Now, we extend our analysis to incorporate anisotropic synchrotron
emission by assuming a toroidal magnetic field configuration. In
Fig. \textbf{\ref{figtong2}}, we interpret the corresponding
intensity maps of the considering BH model under the RIAF model, at
a frequency of $230\mathrm{GHZ}$ with an inclination angle of
$\theta_o=75^\circ$, and a spin parameter $a=0.6$. For quantitative
comparison, the horizontal and vertical intensity profiles are
depicted in Figs. \textbf{\ref{figtong_curve3}} and
\textbf{\ref{figtong_curve4}}. In this perspectives, the entire
morphology remains qualitatively similar to the isotropic case, as
discussed in Fig. \textbf{\ref{figtong1}}, characterized by a
pronounced bright ring encircling by two central dark regions, both
of which slightly reduces with increasing $\ell$. Moreover, as $Q$
increases, the resulting image exhibits a slightly asymmetry,
characterized by enhanced brightness on the side co-rotating with
the BH, due to frame-dragging effects. A significant feature of the
anisotropic case is the observation of a vertically elongated,
elliptical ring structure. This asymmetry appears from the angular
dependence of synchrotron emissivity: photons emitted from the upper
and lower regions of the disk propagate nearly perpendicular to the
magnetic field, resulting in slightly enhanced emission and a
vertically stretched bright ring.

\subsection{BAAF Model}
Now, we discuss the background mechanism of BAAF model, which is
proposed by Hou et al. \cite{bm57,Zhang:2024lsf}. In this framework,
the fluid is assumed to be electrically neutral, with the plasma
fully ionized into electrons and protons. The accreting matter is
constrained to constant-$\theta$ surfaces such as, $u^\theta \equiv
0$. The conservation equation for the mass flow can thus be defined
as
\begin{align}
\frac{d}{dr} \left( \sqrt{-g} \, \rho \, u^r \right) = 0,
\end{align}
with its solution is
\begin{equation}
\rho = \frac{\rho_0}{\sqrt{-g} u^r} \left. \sqrt{-g} u^r
\right|_{r=r_0},
\end{equation}
here, $\rho_0 = \rho(r_0)$ is the mass density at the reference
point, typically taken as $r_0 = r_+$. The projection of the energy
momentum tensor along $u^\alpha$ satisfies
\begin{align}\label{eq:energy_momentum}
d\chi = \frac{\chi + p}{\rho} d\rho,
\end{align}
where $\chi$ denotes the internal energy of the fluid. Further, the
ratio between proton to electron temperature is defined as $U= T_p /
T_e$, and the corresponding internal energy under this approximation
is
\begin{align}\label{eq:internal_energy}
\chi= \rho + \frac{3 (U + 2) \rho m_e \theta_e}{2 m_p},
\end{align}
with $\theta_e = U_B T_e / m_e c^2$ denoting the dimensionless
electron temperature. Using the ideal gas law, the pressure is
\begin{align}\label{eq:pressure}
p = n U_B (T_p + T_e) = \frac{(1 + U) \rho m_e \theta_e}{m_p}.
\end{align}
Putting Eqs.~\eqref{eq:internal_energy} and \eqref{eq:pressure} into
Eq.~\eqref{eq:energy_momentum} and integrating gives
\begin{align}
\theta_e = (\theta_e)_0 \left( \frac{\rho}{\rho_0}
\right)^{\frac{2(1+U)}{3(2+U)}},
\end{align}
in which $(\theta_e)_0$ is the reference temperature at $r_+$. For
computational convenience, we consider that $\rho(r_+,\theta)$ based
on a Gaussian distribution in the $\theta$ direction and, in the
conical solution, take $\theta_e(r_+,\theta)$ to be constant
\begin{align}
\rho(r_+,\theta) = \rho_h \exp\left[-\left( \frac{\sin\theta -
\sin\mu_\theta}{\sigma_\theta} \right)^2 \right], \quad
\theta_e(r_+,\theta) = \theta_h,
\end{align}
here $\mu_\theta$ represents the mean position in the $\theta$
direction and $\sigma_\theta$ is the standard deviation of the
distribution. For M87$^\ast$, observations indicate $\rho_h \simeq
1.5 \times 10^3~\mathrm{g\,cm^{-1}\,s^{-2}}$ and $\Theta_h \simeq
16.86$, corresponding to an electron number density $n_h =
10^6~\mathrm{cm^{-3}}$ and temperature $T_h =
10^{11}~\mathrm{K}$~\cite{Vincent_2022}. For a spherically symmetric
spacetime, the magnetic field configuration simplifies as

\begin{align}
b^\alpha = \frac{\Psi}{\sqrt{-g} u^r} \left[ \left( u_t + \Omega_b
u_\phi \right) u^\mu + \delta_t^\mu + \Omega_b \delta_\phi^\mu
\right],
\end{align}
where $\Psi = F_{\theta \phi}$ is a component of the electromagnetic
field tensor. Here, we consider a separable monopole solution
\begin{align}
    \Psi = \Psi_0 \, \mathrm{sign}(\cos\theta) \, \sin\theta.
\end{align}
The angular velocity of the magnetic field is taken as $\Omega_b =
0.3 \, \Omega_h$, with $\Omega_h = a / (2 r_+)$ indicating the spin
angular velocity of the BH. In this model, we only consider the
anisotropic radiation~\eqref{eq:anisotropic_emissivity}, and the
fluid four velocity is still described by the ballistic
approximation such as, the fluid moves along geodesics.

\begin{figure}[H]
\centering \subfigure[$Q=0.1,
\ell=-0.5$]{\includegraphics[scale=0.45]{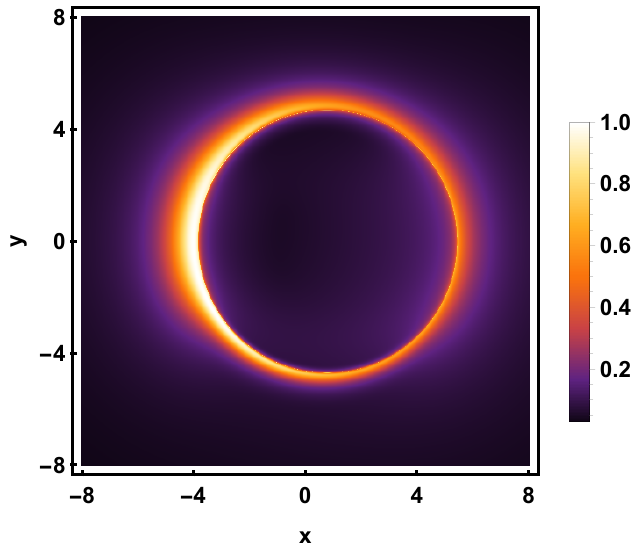}}
\subfigure[$Q=0.1,
\ell=0.1$]{\includegraphics[scale=0.45]{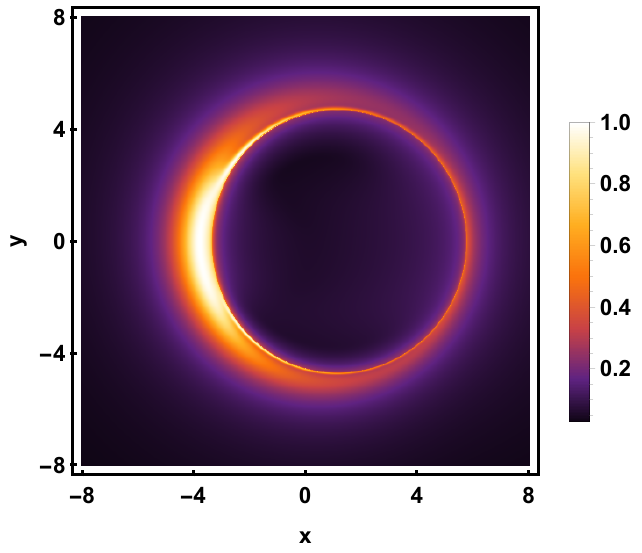}}
\subfigure[$Q=0.1,
\ell=0.5$]{\includegraphics[scale=0.45]{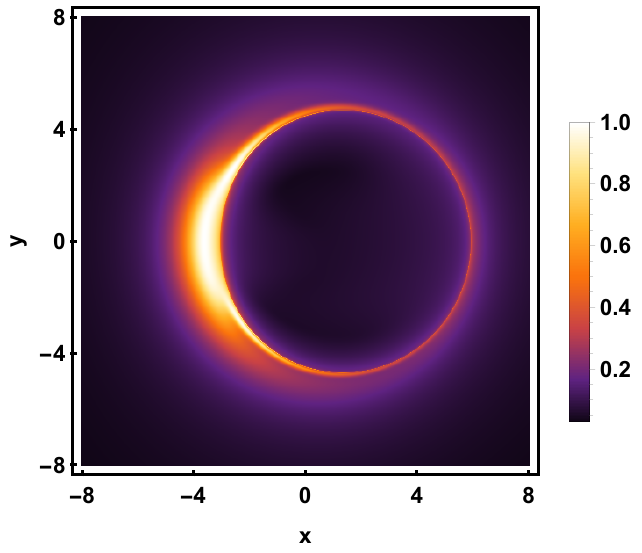}}
\subfigure[$Q=0.3,
\ell=-0.5$]{\includegraphics[scale=0.45]{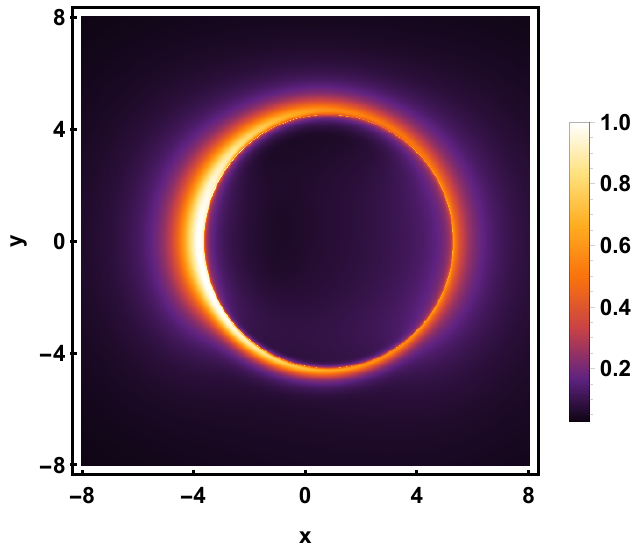}}
\subfigure[$Q=0.3,
\ell=0.1$]{\includegraphics[scale=0.45]{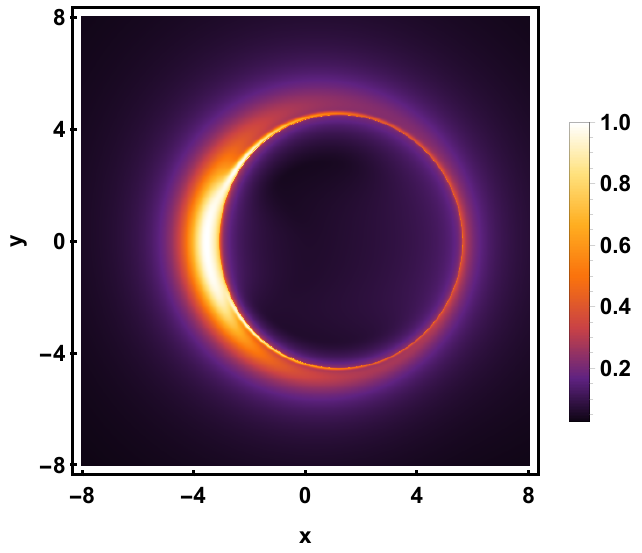}}
\subfigure[$Q=0.3,
\ell=0.5$]{\includegraphics[scale=0.45]{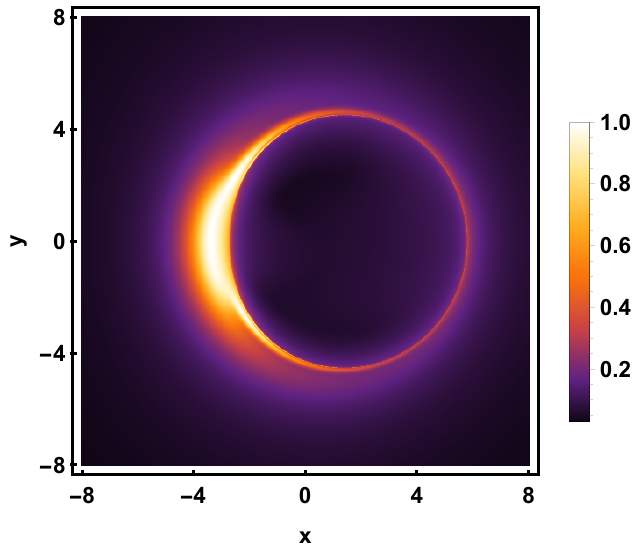}}
\subfigure[$Q=0.5,
\ell=-0.5$]{\includegraphics[scale=0.45]{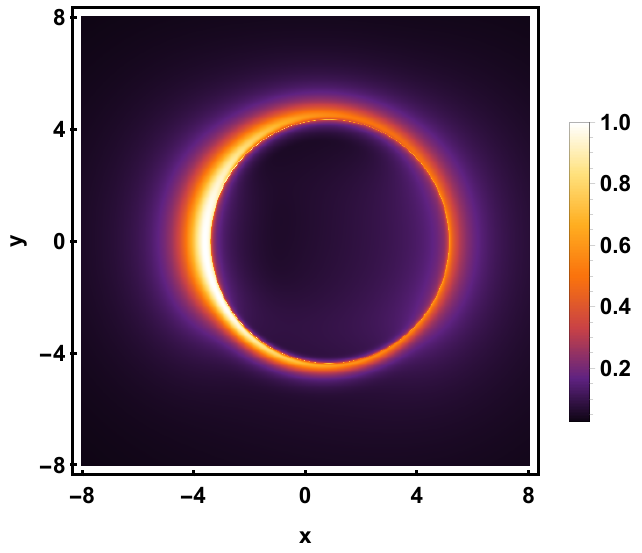}}
\subfigure[$Q=0.5,
\ell=0.1$]{\includegraphics[scale=0.45]{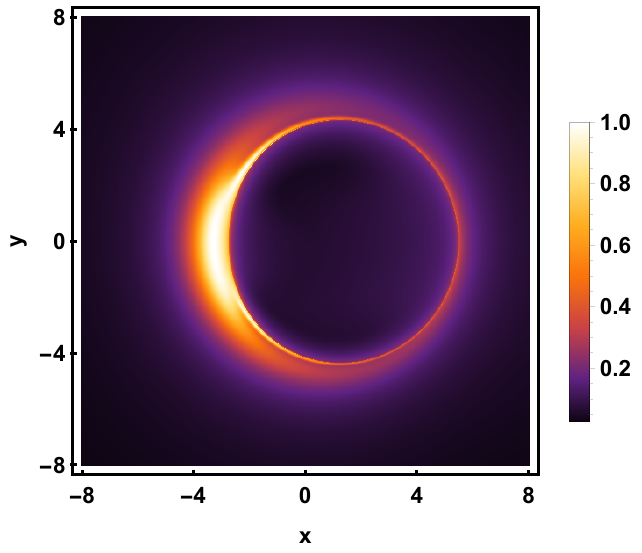}}
\subfigure[$Q=0.5,
\ell=0.5$]{\includegraphics[scale=0.45]{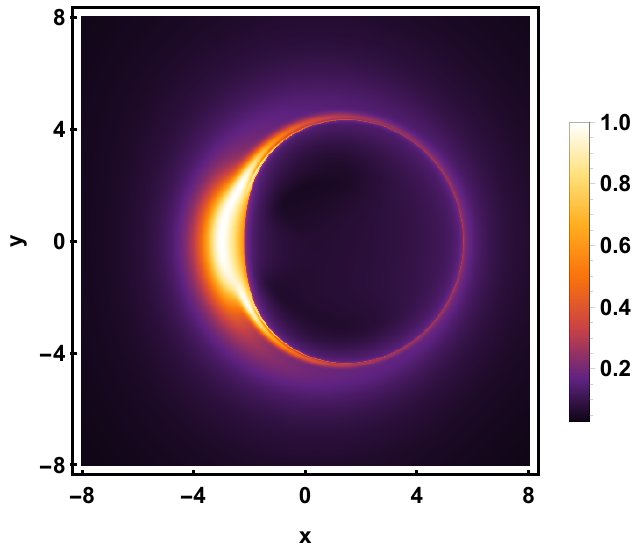}} \caption{Imaging
results of the thick accretion disk for anisotropic radiation under
the BAAF model, and the accretion flow follows the infalling motion.
The observation is performed at a frequency of $230\mathrm{GHz}$
with an inclination angle of $\theta_o=75^\circ$, and a spin
parameter $a=0.6$.}\label{figtong3}
\end{figure}

\begin{figure}[H]\centering \subfigure[Horizontal
direction]{\includegraphics[scale=0.8]{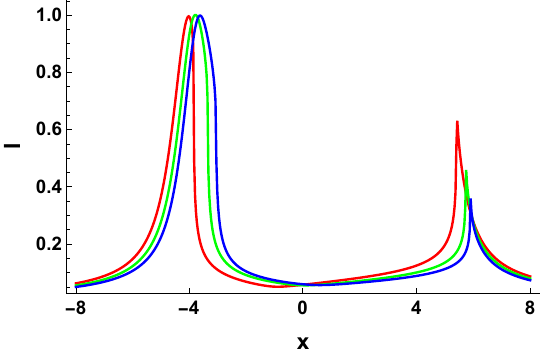}} \subfigure[Vertical
direction]{\includegraphics[scale=0.8]{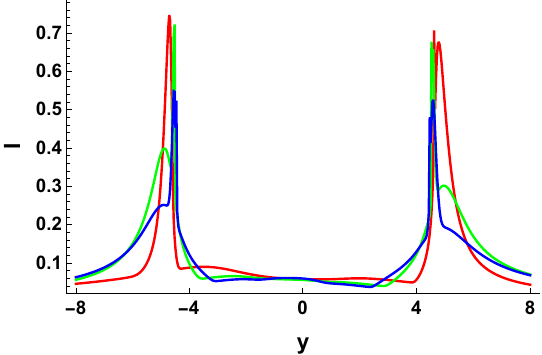}} \caption{Intensity
distribution for anisotropic radiation under the BAAF model, and the
accretion flow follows the infalling motion. The red, green and blue
curves corresponds to $\ell=-0.5,~0.1$ and $0.5$, respectively with
a fixed $a=0.6,~Q=0.1,~\theta_o=75^\circ$.} \label{figtong_curve5}
\end{figure}

\begin{figure}[H]\centering \subfigure[Horizontal
direction]{\includegraphics[scale=0.8]{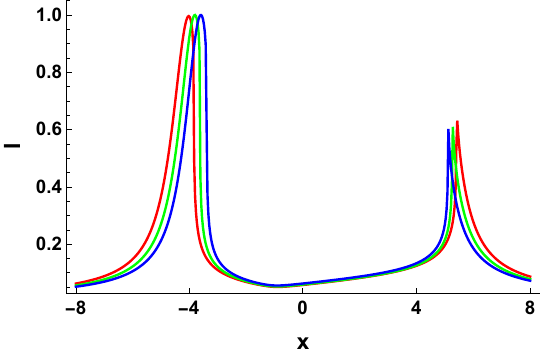}} \subfigure[Vertical
direction]{\includegraphics[scale=0.8]{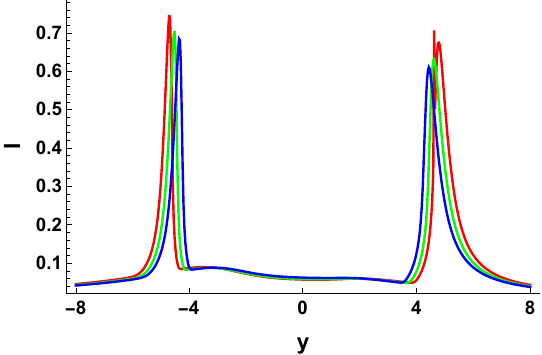}} \caption{Intensity
distribution for anisotropic radiation under the BAAF model, and the
accretion flow follows the infalling motion. The red, green and blue
curves corresponds to $Q=0.1,~0.3$ and $0.5$, respectively with a
fixed $a=0.6,~\ell=-0.5,~\theta_o=75^\circ$.} \label{figtong_curve6}
\end{figure}

\begin{figure}[H]
\centering
\subfigure[$\theta_o=0^\circ$]{\includegraphics[scale=0.45]{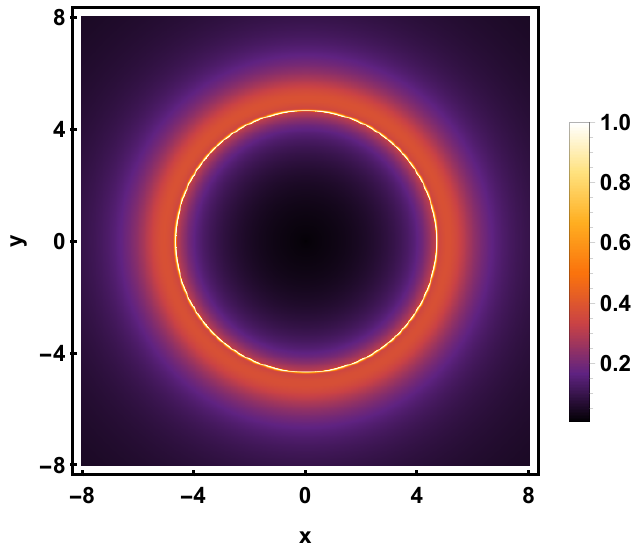}}
\subfigure[$\theta_o=45^\circ$]{\includegraphics[scale=0.45]{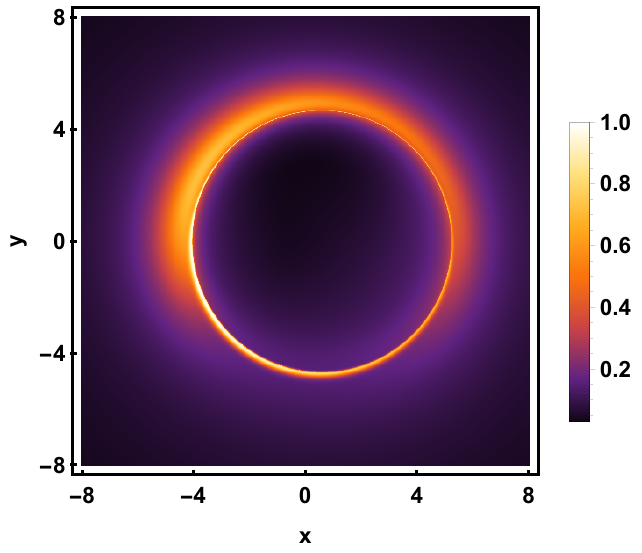}}
\subfigure[$\theta_o=75^\circ$]{\includegraphics[scale=0.45]{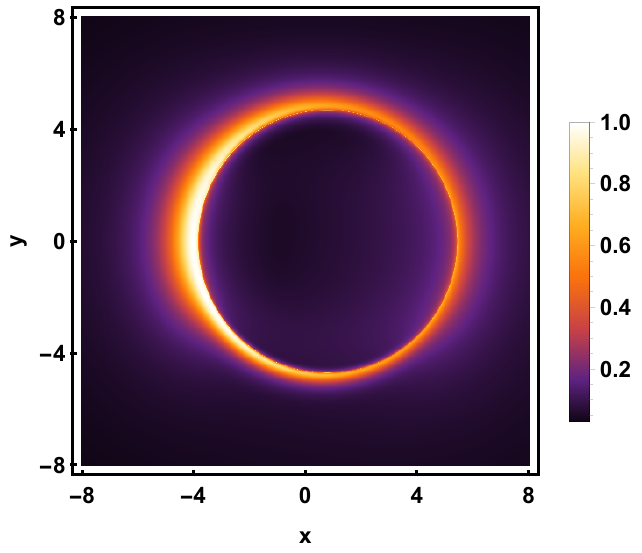}}\caption{Imaging
results of the thick accretion disk for anisotropic radiation under
the BAAF model, and the accretion flow follows the infalling motion.
The observation is performed at a frequency of $230\mathrm{GHz}$
with an inclination angles of $\theta_o=0^\circ,~\theta_o=45^\circ$
and $\theta_o=75^\circ$ from left to right, respectively. For all
cases we fixed $a=0.6,~\ell=-0.5$ and $Q=0.1$.}\label{figtong4}
\end{figure}

\begin{figure}[H]\centering \subfigure[Horizontal
direction]{\includegraphics[scale=0.8]{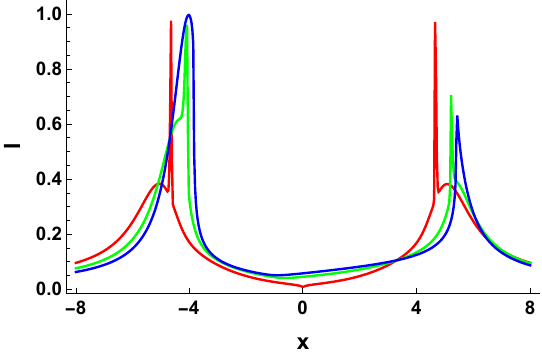}} \subfigure[Vertical
direction]{\includegraphics[scale=0.8]{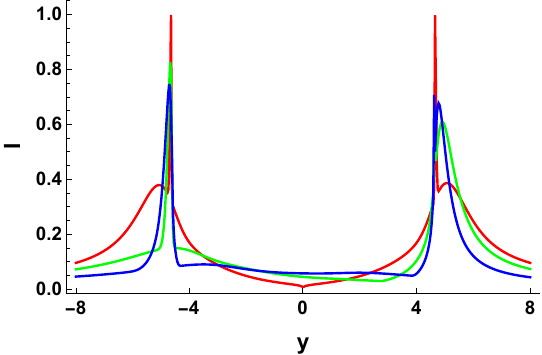}} \caption{Intensity
distribution for anisotropic radiation under the BAAF model, and the
accretion flow follows the infalling motion. The red, green and blue
curves corresponds to $\theta_o=0^\circ,~\theta_o=45^\circ$ and
$\theta_o=75^\circ$, respectively with a fixed
$a=0.6,~\ell=-0.5,~Q=0.1$.} \label{figtong_curve7}
\end{figure}
In Fig. \textbf{\ref{figtong3}}, we exhibit the overall intensity
distribution of the Kerr-Sen-like BH within the fabric of the BAAF
disk model under anisotropic synchrotron emission. The accretion
flow is considered to be purely infalling, and the observation is
performed at a frequency of $230\mathrm{GHZ}$. The corresponding
horizontal and vertical intensity profiles are depicted in Figs.
\textbf{\ref{figtong_curve5}} and \textbf{\ref{figtong_curve6}}.
Overall, the dependence of the image morphology on the BH LSB
parameter $\ell$ and charge $Q$, closely parallels the behavior
observed in the RIAF case. Nevertheless, several notable differences
are observed. Compared with the RIAF framework, as shown in Figs.
\textbf{\ref{figtong1}} and \textbf{\ref{figtong2}}, the bright ring
in the BAAF disk images appears generally thinner, and the
separation between the primary and higher-order images becomes more
pronounced.

We further presents the variation of the intensity distribution with
the alteration of the inclination angles in Fig.
\textbf{\ref{figtong4}}. The associated horizontal and vertical
intensity profiles are displays in Fig.
\textbf{\ref{figtong_curve7}}, which provides a quantitative
comparison of the brightness distributions. For polar viewing, the
bright ring and the dark region lies in the centered. At
$\theta_o=0^\circ$, a clear up-down asymmetry in the bright ring
emerges. When the inclination increases to $\theta_o=45^\circ$, the
dark region slightly emerge inside the ring, with the upper one
significantly thicker, as compared to lower one. At
$\theta_o=75^\circ$, the brightness distribution becomes slightly
nonuniform, with a distinct dark region appearing inside the bright
ring. Moreover, the left-side brightness are more pronounced, and
the bright ring structure are vertically elongated, at large
inclinations. This asymmetry appears from the angular dependence of
synchrotron emissivity: photons radiated from the upper and lower
parts of the disk propagate nearly perpendicular to the magnetic
field, resulting in enhanced emission and a vertically stretched
ring.
\section{Polarization Imaging}
For polarized imaging, we only consider the BAAF model with
anisotropic radiation. Based on the WKB approximation, the
propagation of light satisfies the covariant radiative transfer
equation
\begin{align}\label{eq:covariant_rte}
k^\alpha \nabla_\alpha \mathcal{S}^{\mu\nu} = J^{\mu\nu} +
H^{\mu\nu\alpha\varrho} \mathcal{S}_{\alpha\varrho}.
\end{align}
Here, $k^\alpha$ represents the photon's wave vector,
$\mathcal{S}^{\mu\nu}$ denotes the polarization tensor defining the
polarization state of the light, $J^{\mu\nu}$ describes the emission
properties of the radiation source, and $H^{\mu\nu\alpha\varrho}$
characterizes the response of the propagation medium to the light
ray, including absorption and Faraday's rotation effects
~\cite{Huang:2024bar}. The polarization tensor
$\mathcal{S}^{\mu\nu}$ is directly proportional to the photon's
polarization density matrix. Subsequently, it is Hermitian,
satisfying $\mathcal{S}^{\mu\nu} = \overline{\mathcal{S}}^{\nu\mu}$,
in which the overline denotes complex conjugation, and it is gauge
invariant. Using this gauge invariance, the calculation can be
carried out within a conveniently chosen parallel-transported tetrad
frame The covariant radiative transfer
equation~\eqref{eq:covariant_rte} can then be classified into two
parts. The first part,
\begin{align}
k^\alpha \nabla_\alpha f^\nu = 0, \quad f_\nu k^\nu = 0,
\end{align}
reflects the gravitational effects, where $f^\nu$ is a normalized
spacelike vector orthogonal to $k^\nu$. The second part corresponds
to the radiative transfer along the ray
\begin{align}
    \frac{d S}{d\varpi} = R(\psi) J - R(\psi) M R(-\psi) S,
\end{align}
with the matrices defined as
\begin{align}
    S = \begin{pmatrix}
        \mathcal{I} \\
        \mathcal{Q} \\
        \mathcal{U} \\
        \mathcal{V}
    \end{pmatrix},\qquad
    J &= \frac{1}{\nu^{2}}\begin{pmatrix}
        j_{I} \\
        j_{Q} \\
        j_{U} \\
        j_{V}
    \end{pmatrix},\qquad
    M = \nu\begin{pmatrix}
        a_{I} & a_{Q} & a_{U} & a_{V} \\
        a_{Q} & a_{I} & r_{V} & -r_{U} \\
        a_{U} & -r_{V} & a_{I} & r_{Q} \\
        a_{V} & r_{U} & -r_{Q} & a_{I}
    \end{pmatrix},\nonumber\\
    {R}(\psi) &=
    \begin{pmatrix}
        1 & 0 & 0 & 0 \\
        0 & \cos(2\psi) & -\sin(2\psi) & 0 \\
        0 & \sin(2\psi) & \cos(2\psi) & 0 \\
        0 & 0 & 0 & 1
    \end{pmatrix}.
\end{align}
where $R(\psi)$ is a rotation matrix. The rotation angle $\psi$ is
the angle between the reference vector $f^\alpha$ and the local
magnetic field $b^\alpha$ in the transverse plane of the light ray,
calculated as~\cite{Zhou:2025moa}
\begin{align}
    \psi = \mathrm{sign}(\epsilon_{\alpha\beta\rho\sigma} u^\alpha f^\beta b^\rho k^\sigma)
    \arccos \left( \frac{P^{\alpha\beta} f_\alpha b_\beta}{\sqrt{ (P^{\alpha\beta} f_\alpha f_\beta) (P^{\mu\nu} b_\mu b_\nu) }} \right),
\end{align}
where $P^{\alpha\beta}$ is the induced metric in the transverse
subspace. At the observer, the Stokes parameters are projected onto
the observer's screen, again using a rotation matrix. The
corresponding rotation angle is
\begin{align}
    \psi_0 = \mathrm{sign}(\epsilon_{\mu\nu\alpha\beta} u^\mu f^\nu d^\alpha k^\beta)
    \arccos \left( \frac{P^{\alpha\beta} f_\alpha d_\beta}{\sqrt{ (P^{\alpha\beta} f_\alpha f_\beta) (P^{\mu\nu} d_\mu d_\nu) }} \right),
\end{align}
where $d^\alpha$ is chosen along the $y$-axis of the screen,
$d^\alpha = -\partial_\theta^\alpha$. The projected Stokes
parameters are then
\begin{align}
    \mathcal{I}_o = \mathcal{I}, \quad
    \mathcal{Q}_o = \mathcal{Q} \cos\psi_o - \mathcal{U} \sin\psi_o, \quad
    \mathcal{U}_o = \mathcal{Q} \sin\psi_o + \mathcal{U} \cos\psi_o, \quad
    \mathcal{V}_o = \mathcal{V},
\end{align}
where $\mathcal{I}_o$ represents the intensity. The Stokes
parameters $\mathcal{Q}_o$ and $\mathcal{U}_o$ are associated with
the electric field $\vec{E} = (E_x, E_y)$ by
\begin{align}
    \mathcal{Q}_o = E_x^2 - E_y^2, \quad \mathcal{U}_o = 2 E_x E_y.
\end{align}
Generally, if $\mathcal{U}_o$ is positive, $E_x$ and $E_y$ have the
same sign, and $\vec{E}$ lies in the first or third quadrant; if
$\mathcal{U}_o$ is negative, $E_x$ and $E_y$ have opposite signs,
and $\vec{E}$ lies in the second or fourth quadrant. The sign of
$\mathcal{Q}_o$ indicates whether $\vec{E}$ is aligned closer to the
line $y = x$ or $y = -x$. The parameter $\mathcal{V}_o$
characterizes the circular polarization: positive values indicate
left-handed circular polarization, while negative values correspond
to right-handed circular polarization. From the Stokes parameters,
one can determine both the magnitude and orientation of the
projected linear polarization vector $\vec{f}$ on the observer's
frame. Specifically, its magnitude gives the degree of linear
polarization, whereas its direction defines the electric vector
position angle (EVPA)
\begin{align}
    |\vec{f}| = \mathcal{P}_o = \frac{\sqrt{\mathcal{Q}_o^2 + \mathcal{U}_o^2}}{\mathcal{I}_o}, \qquad
    \arg(\vec{f}) = \Phi_{\text{EVPA}} = \frac{1}{2} \arctan \left( \frac{\mathcal{U}_o}{\mathcal{Q}_o} \right).
\end{align}
Within this mechanism, the Stokes parameters and the linear
polarization vector $\vec{f}$ can be evaluated, enabling a complete
characterization of the polarization characteristics.

\begin{figure}[H]
\centering \subfigure[Stokes parameter
$\mathcal{I}_{o}$]{\includegraphics[scale=0.55]{Po1_1}}
\subfigure[Stokes parameter
$\mathcal{Q}_{o}$]{\includegraphics[scale=0.6]{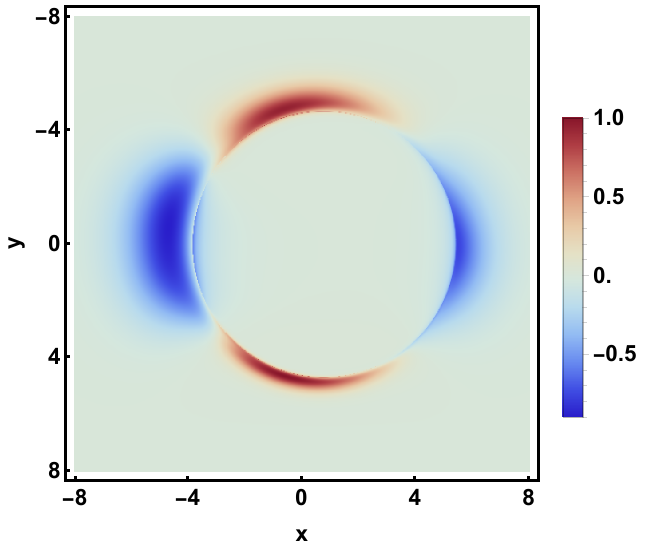}}
\subfigure[Stokes parameter
$\mathcal{U}_{o}$]{\includegraphics[scale=0.6]{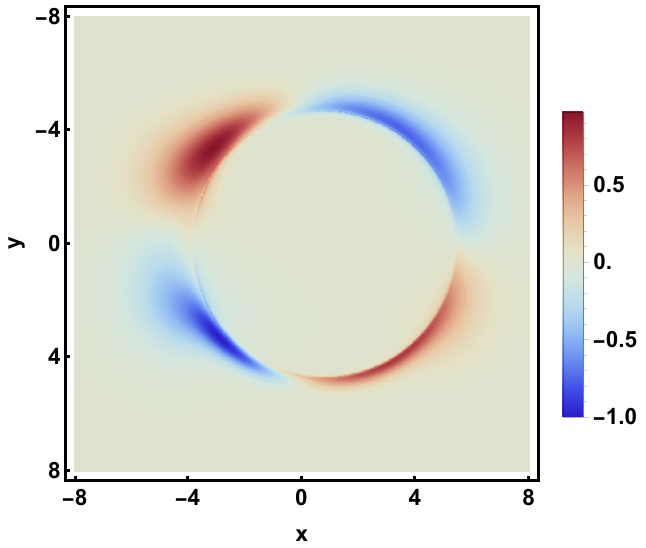}}
\subfigure[Stokes parameter
$\mathcal{V}_{o}$]{\includegraphics[scale=0.6]{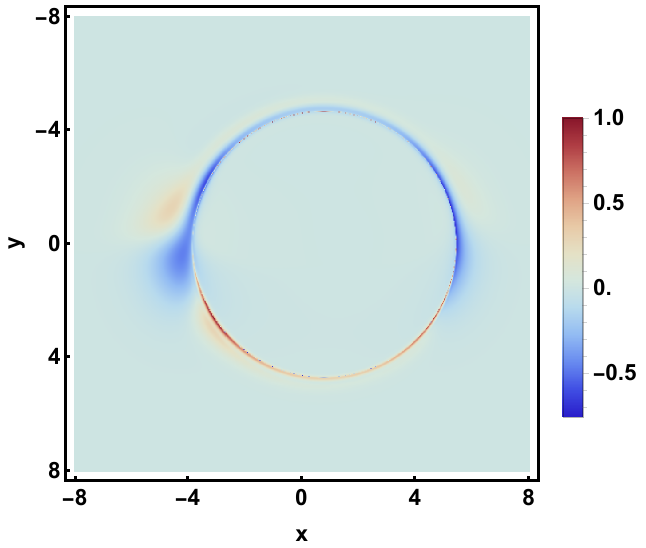}} \caption{The
resulting Stokes parameters $\mathcal{I}_{o}$, $\mathcal{Q}_{o}$,
$\mathcal{U}_{o}$, and $\mathcal{V}_{o}$ under the BAAF disk model
with an infalling motion. The observation is performed at a
frequency of $230\mathrm{GHz}$ with an inclination angle of
$\theta_o=75^\circ$, with a fixed $a=0.6,~\ell=-0.5$ and
$Q=0.1$.}\label{figIQUV}
\end{figure}

\begin{figure}[H]
\centering \subfigure[$Q=0.1,
\ell=-0.5$]{\includegraphics[scale=0.45]{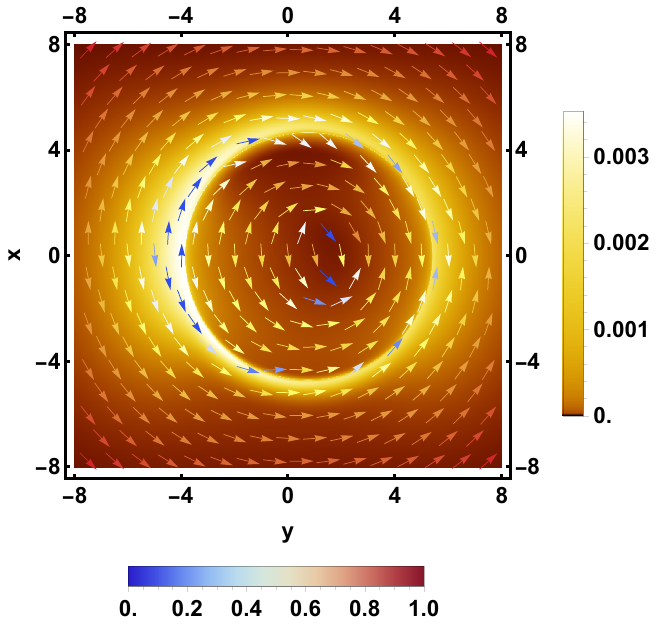}}
\subfigure[$Q=0.1,
\ell=0.1$]{\includegraphics[scale=0.45]{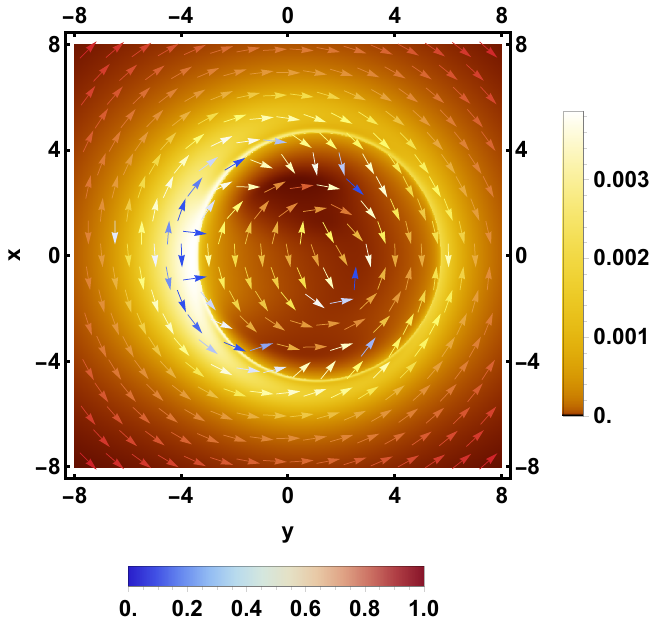}}
\subfigure[$Q=0.1,
\ell=0.5$]{\includegraphics[scale=0.45]{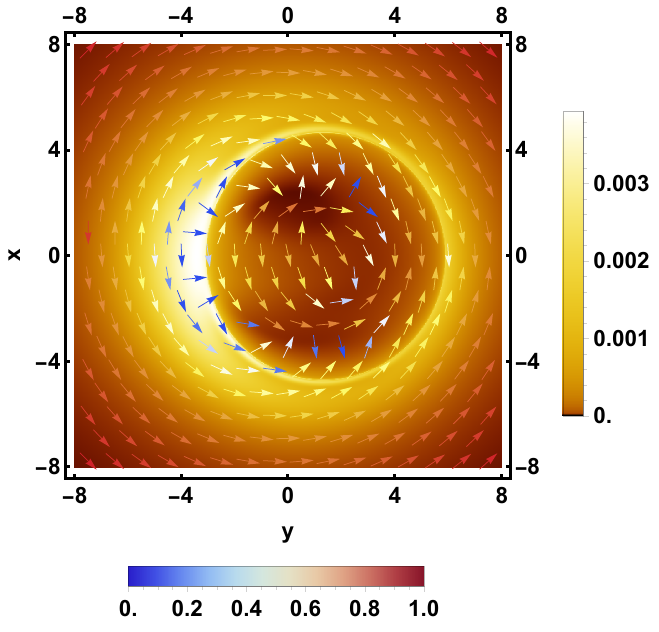}}
\subfigure[$Q=0.3,
\ell=-0.5$]{\includegraphics[scale=0.45]{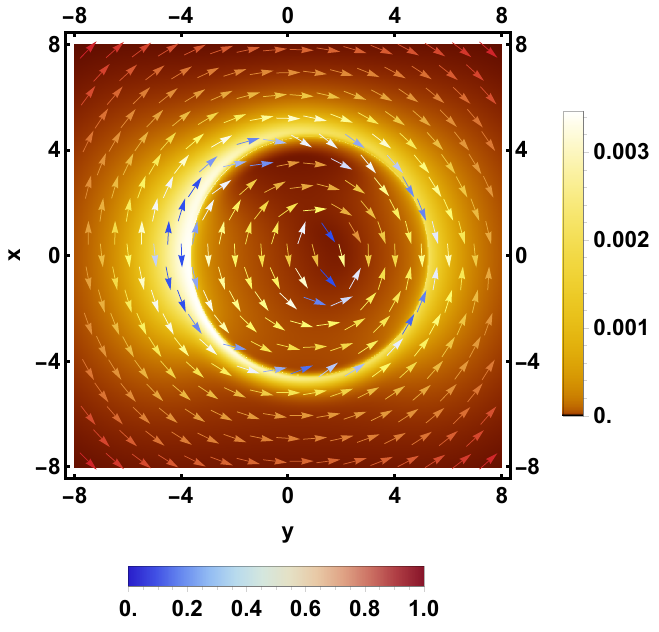}}
\subfigure[$Q=0.3,
\ell=0.1$]{\includegraphics[scale=0.45]{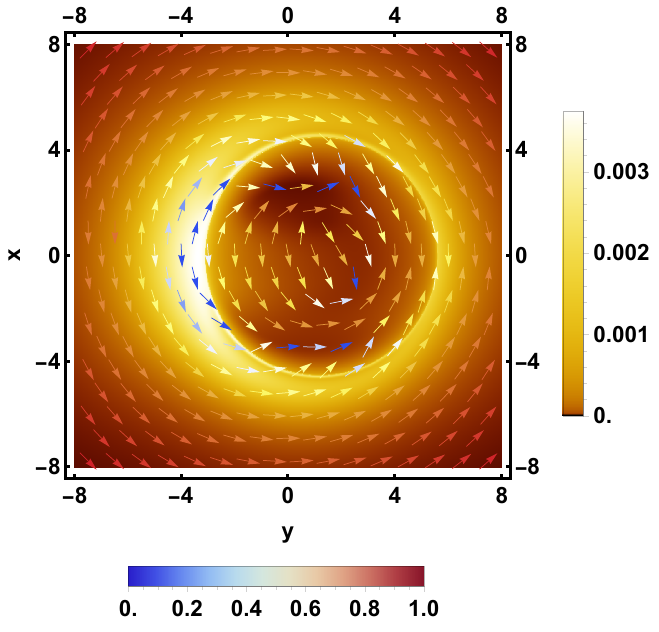}}
\subfigure[$Q=0.3,
\ell=0.5$]{\includegraphics[scale=0.45]{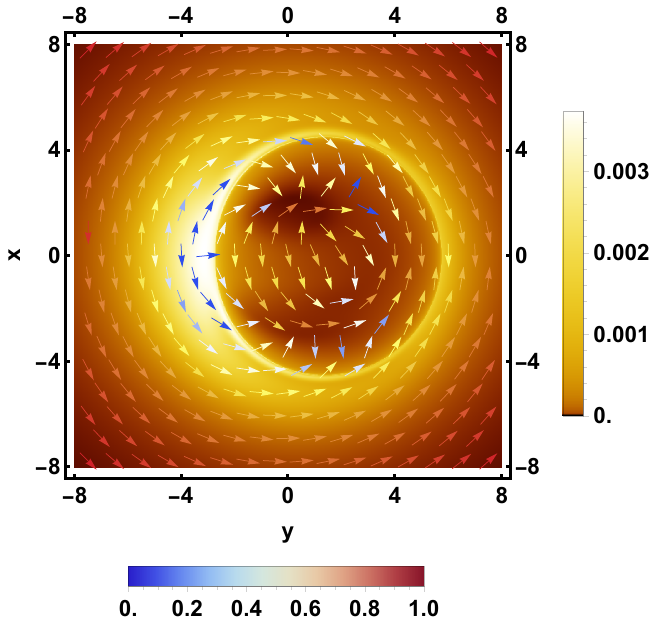}}
\subfigure[$Q=0.5,
\ell=-0.5$]{\includegraphics[scale=0.45]{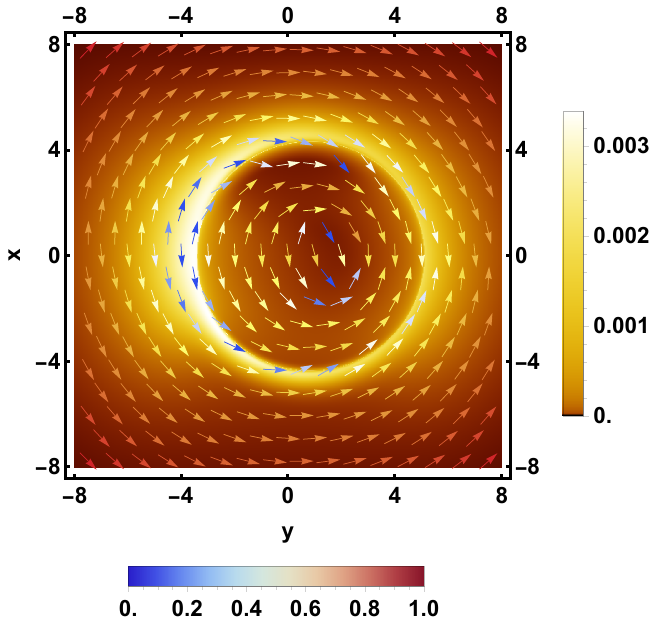}}
\subfigure[$Q=0.5,
\ell=0.1$]{\includegraphics[scale=0.45]{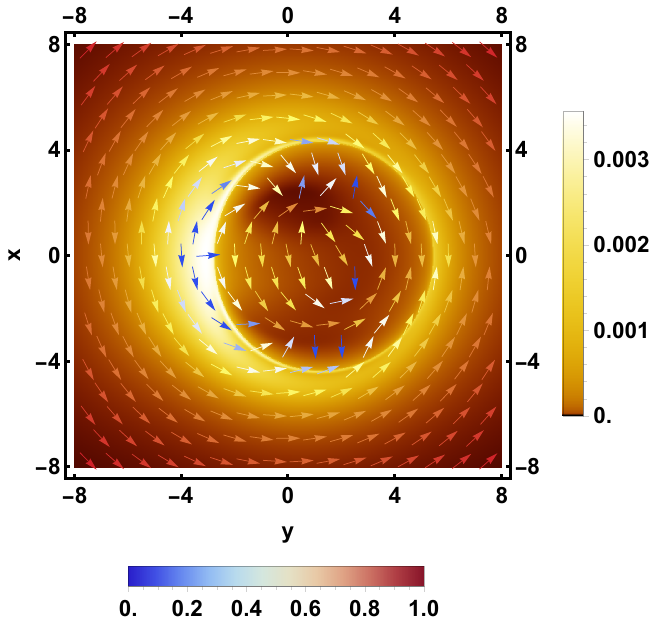}}
\subfigure[$Q=0.5,
\ell=0.5$]{\includegraphics[scale=0.45]{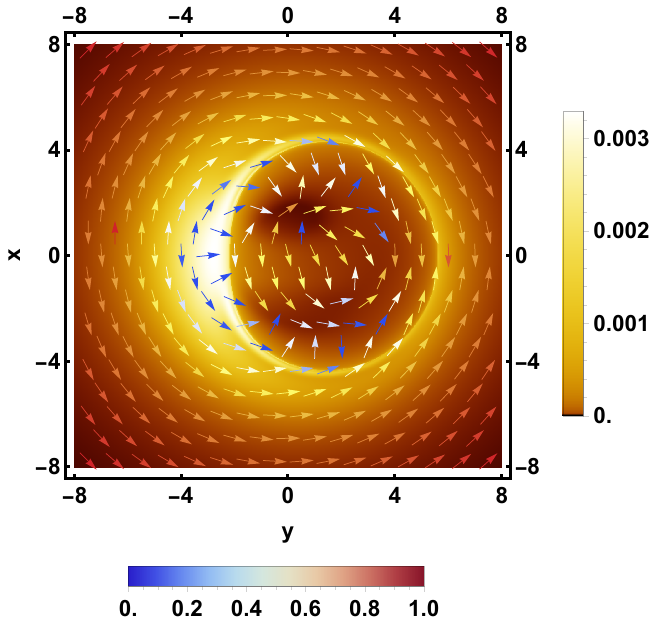}}
\caption{Polarized images of the Kerr-Sen-like BH in the BAAF model
with anisotropic emission. The accretion flow follows the infalling
motion, and the observation is performed at a frequency of
$230\mathrm{GHz}$ with an inclination angle of $\theta_o=75^\circ$,
and a spin parameter $a=0.6$.}\label{figpolar}
\end{figure}
In Fig. \textbf{\ref{figIQUV}}, we presents a representative example
of the observed Stokes parameters $\mathcal{I}_{o}$,
$\mathcal{Q}_{o}$, $\mathcal{U}_{o}$, and $\mathcal{V}_{o}$ under
the BAAF disk model with an infalling motion. The observation is
performed at a frequency of $230\mathrm{GHz}$ with an inclination
angle of $\theta_o=75^\circ$, with a fixed $a=0.6,~\ell=-0.5$ and
$Q=0.1$. The quantity $\mathcal{I}_{o}$ indicates the intensity
distribution. The arrows represents the linear polarization vector
$\vec{f}$, where their color corresponds to the polarization degree
$\mathcal{P}_{o}$, while their orientation specifies the EVPA
$\Phi_{\mathrm{EVPA}}$. Since $\vec{f}$ is always perpendicular to
the magnetic field $\vec{b}$, it can be inferred that the magnetic
field is approximately radial. Combining $\mathcal{Q}_{o}$ and
$\mathcal{U}_{o}$ allows for a qualitative determination of the
direction of the electric vector $\vec{E}$, while $\mathcal{V}_{o} <
0$ indicates right-handed circular polarization. Closer to the BH
event horizon, frame-dragging effects become dominant, causing the
magnetic field lines to twist into a more azimuthal configuration.
Using the flux-freezing condition of ideal magnetohydrodynamics, the
magnetic field remains closely coupled to the plasma motion.
Consequently, the observed rotation of the polarization angle serves
as a direct signature of the frame-dragging influence of the BH on
the accreting plasma. The parameters $\mathcal{Q}_{o}$ and
$\mathcal{U}_{o}$ reach their maxima closer to the higher-order
images and decay rapidly away from this region. The distribution of
$\mathcal{V}_{o}$ indicates right-handed polarization on both sides
of the higher-order images, while the remaining regions displays
left-handed polarization.

Figure \textbf{\ref{figpolar}} presents the impact of $\ell$ and $Q$
on the polarization images. In these images, the bright rings
correspond to the higher-order images, and the dark region inside
originates from the event horizon. The results indicates that as
$\ell$ increases from left to right, the position where the
polarization vectors start to exhibit azimuthal twisting moves away
from the BH. Meanwhile, by examining each column, we find that as
$Q$ increases, the polarization vectors exhibit noticeable changes,
while the corresponding polarized intensity within the bright ring
becomes increasingly prominent. Overall, both $\ell$ and $Q$
significantly influence the polarization properties, reflecting the
impact of the underlying spacetime structure on the observed
polarization behavior.

\section{Summary}
In the present study, we revisited the fundamentals of Bumblebee
gravity, which yields a Kerr-Sen-like BH metric, and analyzed the
visual characteristics surrounded by a geometrically thick accretion
flow. We define the background of null geodesics along with the
definition of photon sphere, and then provide a comprehensive review
of two representative geometrically thick accretion flow models,
which is known as the phenomenological RIAF model and the analytical
BAAF model. We solve the geodesics and the radiative transfer
equations numerically, and observed the corresponding shadow images
and polarization structures with observational frequency at
$230\mathrm{GHz}$.

For the RIAF model, both the isotropic and anisotropic radiation
frameworks are investigated, with the accretion flow mode being the
infalling motion for various values of LSB parameter $\ell$ and
charge $Q$ with an inclination angle of $\theta_o=75^\circ$. The
obtained results show that, in isotropic radiation scenarios,
increasing $Q$ slightly decreases both the size and brightness of
the higher-order image, while increasing $\ell$ alters the shape of
the higher-order image and obscures the horizon's boundary.
Moreover, the image is nearly symmetric in the vertical direction,
although the left-right intensity remains higher than the top-bottom
intensity intensity. Interestingly, at smaller values of $\ell=0.1$,
two dark regions appear inside the higher-order image, with the
upper region slightly darker than the lower one, which is more
obvious with increasing $\ell$. This phenomenon arises from
gravitational lensing effects. We further examined the anisotropic
synchrotron emission by imposing a toroidal magnetic field
configuration. The overall morphology remains qualitatively similar
to the isotropic case, as characterized by a pronounced bright ring
encircled by two central dark regions, both of which slightly reduce
with increasing $\ell$. A significant feature is observed that the
brightness distribution exhibits a slight asymmetry for higher
values of $Q$, characterized by enhanced brightness on the side
co-rotating with the BH, due to frame-dragging effects. The bright
ring appears vertically elongated and slightly elliptical,
reflecting the geometry of the underlying magnetic field
configuration.

We further analyzed the imaging and polarization properties of the
BAAF disk model. For the intensity distribution, we again assume
purely infalling matter with anisotropic synchrotron radiation,
allowing for a direct comparison with the RIAF model. The resulting
intensity maps display a relatively narrower bright ring and a more
pronounced dark central region than those obtained in the RIAF case.
This distinction can be attributed to the fact that, for certain
parameter regimes, the BAAF disk within the conical approximation
becomes geometrically thinner than the corresponding RIAF disk in
some regions. For the polarization images of the BAAF model, both
the LSB parameter $\ell$ and the BH charge $Q$ play a crucial role
in shaping the magnitude and orientation of the polarization
vectors. This suggests that polarization images of the considered BH
model can serve as an effective probe of the underlying spacetime
geometry. Moreover, in contrast to thin-disk scenarios,
gravitational lensing of radiation originating from off-equatorial
regions produce a wider spread of polarization vectors across the
image plane. As a result, thick-disk models exhibit richer and more
intricate polarization signatures.

In conclusion, this work main physical parameters of a BH strongly
influence the intricate properties of its shadow in geometrically
thick, optically thin accretion flows. These configurations provide
a more realistic description of astrophysical environments compared
to idealized thin disk models. A combined analysis of intensity and
polarization images enables a more complete understanding of both
the emitted radiation properties and the surrounding spacetime
geometry. Looking ahead, it would be worthwhile to extend this
framework to other compact objects, such as neutron stars and boson
stars, to identify distinguishing observational features among
different gravitational systems. We hope that these investigations
could provide useful theoretical benchmarks for future high
resolution astronomical observations.\\
{\bf Acknowledgements}\\
This work is supported by the National Natural Science Foundation of
China (Grants Nos. 12375043, 12575069 ), and Chongqing Normal
University Fund Project (Grants No. 26XLB001).\\
Princess Nourah bint Abdulrahman University Researchers Supporting
Project number (PNURSP2026R59), Princess Nourah bint Abdulrahman
University, Riyadh, Saudi Arabia.

\end{document}